\documentstyle[tables]{lamuphys}
\def\tr{{\rm Tr}}
\begin{document}
\title{{Light-Front QCD and Heavy Quark Systems}\thanks{Lecture 
Notes, based on five lectures given at ``The First International 
School on Light-Front Quantization and Nonperturbative Dynamics 
--- Theory of Hadrons and Light-Front QCD", IITAP, Ames, IA, USA,
May 1996.}}
\author{{\bf Wei-Min Zhang}\thanks{E-mail address: 
wzhang@phys.sinica.edu.tw}}
\institute{Institute of Physics, Academia Sinica, Taipei 11529; 
Department of Physics, National Tsing Hua University, Hsinchu 30043, 
Taiwan, ROC}

\maketitle

\begin{abstract}
In this series of lectures, I shall begin with the current 
investigations on phenomenology of hadron dynamics 
to demonstrate the importance of solving hadronic bound 
states within the framework of light-front (LF) QCD. Then, I 
will describe the basic procedure how to formulate the 
canonical theory of LFQCD, including light-front quantization 
of QCD, light-front gauge singularity, and light-front 
two-component formalism. I will 
also present a complete one-loop QCD calculation in terms 
of the light-front time-ordering perturbation theory, in 
comparison with the usual covariant perturbative QCD calculation. 
Following thereby I will discuss the development of heavy-quark 
effective theory and the manifestation of heavy quark symmetry on 
the light-front. Finally, by applying recently developed similarity 
renormalization group approach to light-front heavy quark 
effective theory, I will show a rigorous derivation of quark 
confinement interaction from   LFQCD and its application
to solve heavy hadron bound states. 
\end{abstract}

\section{Hadronic Phenomenology in the LF Formulation}

\subsection{An Overview}

Simply speaking, the main task in the investigation of hadronic 
physics is how to provide a QCD description of hadronic
structure. More specifically, how can we compute directly 
from QCD the fruitful hadronic properties, such as the hadronic 
structure functions in lepton-nucleon deep inelastic scatterings, 
the partonic fragmentation functions in high energy hadron-hadron
or $e^+e^-$ collisions, and many hadronic form factors in various 
hadronic decay processes.  However, although 
QCD has been accepted as a fundamental theory of the strong 
interaction that governs the underlying dynamics of hadronic
constituents, a complete QCD description to
hadronic structure is still lacking.  In this series of lectures,
I will attempt to show you that the light-front formulation of field 
theory may provide a natural and systematic QCD description 
to all the processes mentioned above  [\cite{zhang94}]. 

Historically, light-front dynamics played a very important role
in every step of the development of the strong 
interaction theory. The most important application of light-front 
dynamics to hadronic physics is perhaps the parton phenomena in 
the lepton-nucleon deep inelastic scatterings (DIS). As it is 
well-known, DIS probes hadronic dynamics near the light-cone. 
Physically, the DIS phenomena can be understood in terms of 
Feynman's parton picture  [\cite{Feynman72}]. While, only the light-front 
formulation of field theory can provide a natural quantum field theory 
description of parton dynamics. For examples, the leading contributions 
of the unpolarized 
structure function $F_2(x,Q^2)$ and the polarized structure 
function $g_1(x,Q^2)$ in DIS are simply written in terms of 
hadronic matrix elements on the light-front surface $\xi^-=0$:
\begin{eqnarray}
	&& {F_2(x,Q^2)\over x} = {1\over 2\pi P^+} \int d\eta
		e^{-i\eta x} \langle ps | \psi_+^\dagger(\xi^-)
		{\cal Q}^2 \psi_+(0) - h.c. | ps \rangle , \label{F2} \\
	&& g_1(x,Q^2) = {1\over 4 \pi S^+} \int d\eta
		e^{-i\eta x} \langle ps | \psi_+^\dagger(\xi^-)
		\gamma_5 {\cal Q}^2 \psi_+(0) + h.c | ps \rangle . 
		\label{g1}
\end{eqnarray}
Here, $\psi_+(\xi)={1\over 2}\gamma^0\gamma^+ \psi(\xi)$ is the 
light-front quark field operator,  $\eta = {1\over 2}p^+\xi^-$,
$\xi^- = \xi^0 - \xi^3$, ${\cal Q}$ the quark charge operator, and
$|ps\rangle$ the hadronic states. 
It can be shown that in the light-front field theory, Eqs.(\ref{F2})
and (\ref{g1}) are proportional to the momentum and 
helicity distributions of partons (quarks and gluons) inside 
hadrons respectively.  Other structure functions ($F_L$ 
and $g_2$) also have a similar but a bit complicated expressions. 
Nevertheless, it is obvious that if we knew the hadronic bound 
states $|ps\rangle$ from QCD on the light-front, we could completely 
understand the QCD dynamics of DIS.

Another measurement of hadronic structure in terms of light-front 
hadronic matrix elements is the parton fragmentation functions 
in hadron-hadron and other collisions. During high-energy
collisions, many hard partons are produced and then
are hadronized.  Hadronization processes can be characterized 
by the so-called fragmentation functions which is also introduced
initially by Feynman  [\cite{Feynman72}]. Physically, quark and 
gluon fragmentation functions are probabilities of finding  
hadrons in a hard parton produced in collisions. 
These fragmentation functions can be defined as matrix elements of 
quark and gluon operators at light-front separations. For examples, 
the unrenormalized quark fragmentation function is given by 
 [\cite{Collins82}]
\begin{equation}
	f_{A/q}(z) = {z\over 18\pi} \int dx^- e^{-ip^+x^-/z}
		\tr\langle 0| \psi_+(0) |ps\rangle \langle
		ps| \psi_+^\dagger(x^-) |0\rangle,
\end{equation}
and the gluon fragmentation function is defined as
\begin{equation}
	f_{A/g}(z) = {-z\over 36\pi k^+} \int dx^- e^{-ik^+x^-/z}
		\langle 0| F^{+\mu}(0) |ps\rangle \langle
		ps| F^+_\mu (x^-) |0\rangle.
\end{equation}
where Tr traces the color and Dirac components of quarks. 
 Again, if we knew hadronic 
bound states from QCD on the light-front, we could directly 
study the QCD dynamics represented by these fragmentation functions.

In recent years, light-front formulation has also been widely 
used in the phenomenological study of hadronic form factors involving 
in various hadron elastic scatterings and decay processes, by the 
use of the so-called relativistic quark model or light-front quark 
model  [\cite{Tev77}]. Simply speaking, light-front quark model is based 
on truncated Fock space expansion of light-front bound states (upon 
only the valence quark states) and then phenomenologically determines 
the valence Fock states' amplitude (the wavefunction). Unlike the 
study of the structure functions and the fragmentation functions where 
the use of light-front description can make the physical picture 
manifestation, the interesting feature of using light-front 
description to hadronic decay processes is that the simple boost 
operations and the transparent relativistic properties containing 
in light-front bound states may allow one to describe hadronic 
form factors for entire kinematic range of momentum transfer for
these space-like processes. This is quite different from descriptions 
of other hadronic 
quark models, such as the nonrelativistic constituent quark model and 
the beg model, which are normally believed to be applicable only for the 
processes involving small momentum transfer. Very recently, applications 
of light-front quark model have also been extended to the description of 
various heavy meson decay processes, although most of the investigations 
are limited to the calculations of form factors at zero momentum 
transfer, due to the limitation of using the light-front quark model for 
time-like processes. Extending the light-front quark model incorporated 
with higher Fock space contribution (a more realistic light-front 
bound state description) may make the description of hadronic decay 
form factors become possible for the entire kinematic range of momentum 
transfer.  Nevertheless, all the hadronic form factors are extracted
from some hadronic matrix elements, such as,
\begin{equation}
	\langle H'(p') | \Gamma | H(p) \rangle,
\end{equation}
where $\Gamma$ is a transition operator in the corresponding 
process.  Again, if we knew the associated hadronic bound 
states that solved from QCD on the light-front, we would have a 
true QCD description of hadronic decay processes.

The above analysis indicates that once we know how to solve the 
hadronic bound states from QCD, especially for these defined on
a surface of light-front, we can directly calculate various 
hadronic matrix elements involved in many hadronic processes.  
Then a true QCD description of hadronic physics may be realized. 
This series of lectures is devoted to the light-front formulation
of QCD dynamics and the attempt of solving hadronic bound states, 
especially the heavy hadron bound states, directly from such a 
formulation. In the first lecture, I will mainly discuss the 
general structure of hadronic bound states on the light-front.

\subsection{General Structure of Light-Front Bound States}

In the standard language of field theory, relativistic
bound states and resonances are identified by the 
occurrence of poles in Green functions. Although the 
information extracted from this approach provides a 
good definition of physical particles, the ordinary 
wave function structure of bound states in the usual 
quantum mechanics language is lacked. As a
result,  wave function amplitudes extracted from
Green functions may not be universally valid in the calculations 
of various hadronic matrix elements that measured
in experiments. In order to understand  hadronic
structure in terms of hadronic bound state wavefunctions
(which is the most transparent picture in quantum theory), 
the explicit form of  hadronic bound states on some 
fixed time surface is wanted.

However, solving bound states in field theory
as an eigenstate problem has not been well established.  
One may define the bound states as eigenstates of 
$P^0$ and determine these states by solving the 
eigenequation of $P^0$.  But $P^0$ is a square root function 
of the momentum and mass operators which does not give us 
a clear picture of the Schr\"{o}dinger's eigenstate equation 
in quantum mechanics. The widely used framework of finding 
relativistic bound states is the Bethe-Salpeter equation.  
However, Bethe-Salpeter equation itself involve many 
unsolved problems, such as the physical interpretation of 
the Bethe-Salpeter amplitudes, and the numerical difficulty 
in solving the Bethe-Salpeter equation in space-time space, etc. 
Some approximations, such as instant-time approximation, may 
simplify the Bethe-Salpeter equation. But with such 
approximation, the main properties of relativistic dynamics, 
namely the boost dynamics, will be lacking. In other words, 
the results may be no longer relativistic.

Also, in principle, a relativistic bound state can always 
be written as an operator function of the particle creation 
operators acting on the vacuum of the theory.  However, for 
many theories that we are interested in, especially for QCD, 
the vacuum is very complicated.  With a complicated vacuum,
formally writing down a relativistic bound state as a series
of Fock space expansion also becomes very difficult.

However, these subtle problems may be removed when we look at 
the bound states on the light-front.

i). {\it Light-Front Vacuum}.
In the equal-time framework, the vacuum of QCD is crucial 
for a realization of chiral symmetry breaking and color 
confinement. It is also a starting point in the construction 
of hadronic bound states.  However, the understanding of 
the true QCD vacuum is still very limited, although a lot of 
informative work has been carried out in the past two decades 
based on the instanton phenomena  [\cite{thooft76}] and the 
QCD sum rule  [\cite{QCDsum79}].

In the light-front coordinates, a particle's momentum is
divided into the longitudinal component and the transverse 
components. For a physical (on-mass-shell) particle, its 
longitudinal momentum, $k^+=k^0+k^3$, cannot be negative
since the energy of a physical state always dominates its momentum.
As a result, the light-front vacuum for any interacting field
theory can only be occupied by the particles with zero-longitudinal
momentum, namely
\begin{equation}
  |vac \rangle_{LF} = f(a^{\dagger}_{k^+=0}) | 0 \rangle \, ,
\end{equation}
so that $P^+|vac\rangle_{LF}=0$, where $P^+=\sum_ik_i^+$. At this 
point, the light-front vacuum is still not simple. In the past several 
years, to obtain a nontrivial light-front vacuum, many tried to solve 
the so-called zero-mode (the particles with $k^+=0$) problem [\cite{Burhkart}].

To construct hadronic bound states consisting of many quarks
and gluons, one will naturally ask whether it is possible to
express hadronic states in terms of Fock space expansion 
with a trivial vacuum.  It is obvious that if we could ``remove'' 
from the theory the basic constituents with zero longitudinal 
light-front momentum, the vacuum of the full interacting theory 
would be the same as the free field theory, namely
\begin{equation}
  |vac\rangle_{LF} = |0\rangle \, .
\end{equation}

It must note that here ``removing'' from the theory the basic 
constituents 
with zero longitudinal light-front momentum does not mean to 
simply ignore  dynamics of these constituents and their 
contributions to the bound states. Mathematically, one can
remove these constituents with zero longitudinal momentum by 
either using a prescription that requires the field variables
to satisfy the antisymmetric boundary condition in the light-front
longitudinal direction  [\cite{zhang93a}] or dealing with a 
cutoff theory that imposing a cutoff, $k^+ \geq \epsilon$, 
on the momentum expansion of each field variable, where 
$\epsilon$ is a small number  [\cite{Wilson94}]. Thus, the 
positivity of longitudinal momentum with such a prescription
or an explicit cutoff ensures that the light-front vacuum must
be trivial. Now a relativistic bound state 
can be expressed as an ordinary Fock state expansion:
\begin{equation}
  |\Psi \rangle = f(a^\dagger, b^\dagger, d^\dagger) | 0 \rangle \, .
\end{equation}
For QCD, $a^\dagger, b^\dagger$ and $d^\dagger$ are the gluon,
quark and antiquark creation operators with nonzero longitudinal 
momentum, and $f(a^{\dagger}, b^\dagger, d^\dagger)$ must also 
be a color singlet operator as a polynomial function of 
$\{~a^\dagger, b^\dagger, d^\dagger ~\}$.

ii). {\it Light-Front Bound State Equation}.
Once the light-front vacuum becomes trivial and the light-front
bound states for various hadrons are expanded in terms of 
the Fock space, the dynamic equation to determine these states
is rather simple. Explicitly, a hadronic bound state labeled 
by $\alpha$ with total longitudinal and transverse 
momenta $P^+$ and $P_{\bot}$, and helicity (the total spin 
along the longitudinal direction) $\lambda$ can be
expressed as follows:
\begin{equation}
  |\alpha ,P^+,P_{\bot},\lambda \rangle = \sum_{n,\lambda_i}
  \int' \frac{\,dx_id^2k_{\bot i}\,}{2(2\pi)^3}
  |n,x_iP^+,x_iP_{\bot}+k_{\bot i},\lambda_i \rangle
  \Phi_{n/\alpha}(x_i,k_{\bot i},\lambda_i)\, , \label{wf1}
\end{equation}
In Eq.(\ref{wf1}), $n$ represents $n$ constituents contained 
in the state $|n,x_iP^+,x_iP_{\bot}+k_{\bot i},\lambda_i\rangle$, 
$\lambda_i$ is the helicity of the $i$-th constituent, and 
$\int'$ denotes the integral over the space:
\begin{equation}
  \sum_i x_i =1\, , \quad \quad \mbox{and} \quad \quad
  \sum_i k_{\bot i} = 0 \, ,
\end{equation}
where $x_i$ is the fraction of the total longitudinal momentum that the
$i$-th constituent carries, and $k_{\bot i}$ is its relative transverse
momentum with respect to the center of mass frame:
\begin{equation}
  x_i = \frac{\,p_i^+\,}{\,P^+\,} \, , \quad \quad
  k_{i\bot}=p_{i\bot}-x_i P_{\bot} \, ,
\end{equation}
with $p_i^+$, $p_{i\bot}$ being the transverse and longitudinal momentum
of the $i$-th constituent.  $\Phi_{n/\alpha}(x_i,k_{\bot i},\lambda_i)$ 
is the amplitude of the Fock state $|n,x_iP^+,x_iP_{\bot}+k_{\bot i},
\lambda_i \rangle$ which satisfies the following normalization condition:
\begin{equation}
  \sum_{n,\lambda_i} \int' \frac{\,dx_id^2k_{\bot i}\,}{2(2\pi)^3}
  |\Phi_{n/\alpha}(x_i,k_{\bot i},\lambda_i)|^2 = 1 \, .
\end{equation}

The eigenstate equation that the wave functions obey on the
light-front is obtained from the operator Einstein equation
$P^2=P^+P^--P_{\bot}^2=M^2$:
\begin{equation}
  H_{LF}|\alpha ,P^+,P_{\bot},\lambda \rangle =
  \frac{\,P_{\bot}^2+M_{\alpha}^2\,}{P^+}|\alpha ,P^+,P_{\bot},\lambda
  \rangle \, ,
\end{equation}
where $H_{LF}=P^-$ is the light-front Hamiltonian. Futhermore, 
since the boost on the light-front only depends on kinematics, 
boosting a bound state from one Lorentz frame to any other frame is 
quite simple, and is dynamically independent  [\cite{zhang94}].  
Thus, if we found the bound state in the rest frame, we could 
completely understand the particle structure in any frame. 
This is not true in the instant form. In the instant form, the
solutions in the rest frame are not easily boosted to other Lorentz
frames due to the dynamical dependence of the boost transformation.
Therefore, in each different Lorentz frame, one needs to solve the
bound state equation of $P^0$ to obtain the corresponding wave
functions.  This is perhaps the reason why one has not established
a reliable approach to construct relativistic wave functions in the
instant field theory in terms of the Schr\"{o}dinger picture. This
obstacle is  removed on the light-front.

To see the explicit form of the light-front bound state equation,
let us consider a meson wave function (for instance, a pion).
The light-front bound state equation can be expressed as:
\begin{equation}
\begin{array}{l}
  \biggl ( m_{\pi}^2- \sum_i \frac{\,k_{i\bot}^2+m_i^2\,}{x_i}\biggr )
  \left (\begin{array}{c} \Psi_{q\bar{q}} \\ \Psi_{q\bar{q}g} \\
  \vdots \end{array} \right ) \\ 
  \hspace{0.8cm} = \left ( \begin{array}{ccc}
  \langle q\bar{q}|H_{int}|q\bar{q} \rangle & \langle q\bar{q}
  |H_{int}|q\bar{q}g\rangle & \cdots \\ \langle q\bar{q}g|H_{int}
  |q\bar{q} \rangle & \cdots & \\ \vdots & & \end{array} \right ) 
  \left (\begin{array}{c}\Psi_{q\bar{q}} \\ \Psi_{q\bar{q}g} \\
  \vdots \end{array} \right ) \, . \end{array} \label{bse}
\end{equation}
Of course, to exactly solve the above equation for the whole Fock 
space is still impossible.  Practically, one has to truncate the 
Fock space to only include these Fock states with a small number of 
particles. For example, one may truncate all the high order Fock space 
sectors (approximately) from the valence constituent space.  Then
the light-front bound state equation is reduced to the light-front
Bethe-Salpeter equation:
\begin{equation} \label{lfbse}
  \biggl ( m_{\pi}^2- \frac{\,k_{\bot}^2+m^2\,}{\,x(1-x)\,}\biggr )
  \Psi_{q\bar{q}}(x,k_{\bot})= \int \frac{\,dyd^2k'_{\bot}\,}{2(2\pi)^3}
  V_{eff}(x,k_{\bot},y,k'_{\bot})\Psi_{q\bar{q}}(y,k'_{\bot}) \, .
\end{equation}
Note that in Eq.(\ref{lfbse}), $V_{eff}$ denotes an effective 
two-body interaction kernel. In other words, by ``truncating"
the Fock space to only keep the valence quark states, the 
complicated Eq.(\ref{bse}) is reduced to the manable Eq.(\ref{lfbse}) 
but the dominant contribution of higher Fock space to the bound 
states must be now described effectively by $V_{eff}$. The residual
effect should be manageable in the framework of perturbation theory.
A true nonperturbative QCD solution to the hadronic bound states
is if one were able to derive these effective interactions 
directly from QCD rather than that phenomenologically are put 
by hand. This will be discussed in the last Lecture.

\subsection{Phenomenological Hadronic Bound States on the 
Light-Front}

At the present time, how to solve for the bound states discussed
above from QCD is still unclear.  Hence, it may be useful to 
have some insights into the light-front behavior of the meson and 
baryon wave functions which have been constructed phenomenologically 
in describing hadrons.
In fact, the phenomenological light-front meson and baryon bound
states have been studied extensively in the last few years, based
on the light-front quark model or light-front wavefunction 
description. The motivation of light-front quark model is to provide 
a simple relativistic constituent quark model for mesons and baryons 
that can yield a consistent description of the hadronic processes
for both low and high momentum transfer.

The general construction of the phenomenological wave functions
is motivated by that of the non-relativistic constituent quark model.  
The constituent quark model has been very successful in the description
of hadronic spectroscopy with a very simple structure, namely
that all mesons consist of a quark and antiquark pair and the baryons
are made of three constituent quarks, their wave functions satisfy
the $SU(6)$ classification and Zweig's rule which suppresses
particle production in favor of rearrangement of constituents
for hadrons  [\cite{close79}]. However,
such a simple picture is very difficult to be understood within
QCD, due to its nonrelativistic assumption and due to our belief
that  QCD vacuum must be very complicated so that hadrons 
must contain an infinite number of quark-antiquark pairs and 
gluons.

Light-front bound states describe the relativistic
hadronic structure with a nonrelativistic form.  Furthermore, 
the simple vacuum state on the light-front ensures the validity 
of the Fock state expansion of hadronic states.  With the assumption 
of existence of constituent quarks (of masses of hundreds of 
MeVs), the leading approximation to hadronic states that consist of a
quark-antiquark pair for mesons and a three-quark cluster for 
baryons should be a reasonable starting point. More theoretical 
discussion for such a assumption from low energy QCD will be 
given later.

However, it must note that there is a subtle problem in the description
of hadronic structure in terms light-front bound states. That is,
it is not easy to identify the light-front hadronic bound states
with hadronic states which are commonly characterized by spin as
a good quantum number. On the light-front, we are unable to 
kinematically construct the hadronic bound states with fixed spin.
The light-front bound states discussed in the last section are
labeled by helicity rather than spin. In these calculations of the
parton distribution and fragmentation functions, the hadronic 
bound states are defined or classified in terms of the helicity. 
However, when we use the light-front bound states to compute the 
hadronic structural quantities, such as hadronic decay form factors 
and coupling constants, we must have states with a definite spin. 
A general solution to the spin problem on the light-front has not 
been found. However, phenomenologically, the helicity part of the 
bound states on the light-front can be transformed to a light-front 
spin part via the so-called Melosh transformation (which is exact 
{\it only} for free quark theory) such that the hadronic states may be 
projected (approximately but no necessary to be correct) from the 
set of light-front bound states labeled with helicities.
Here, I list some meson and baryon light-front bound states that 
have been used to calculate various hadronic quantities in the 
past few years.

The general form of the phenomenological light-front hadronic
bound states has a similar structure to the constituent quark
model states: for meson states (with only the $q\bar{q}$
Fock space sector),
\begin{equation}
  |P^+,P_{\bot},SS_3\rangle 
   = \int \frac{\,dxd^2k_{\bot}\,}{16\pi^3}
  \sum_{\lambda_1\lambda_2}\Psi_m^{SS_3}(x,k_{\bot},\lambda_1,
  \lambda_2)|x,k_{\bot},\lambda_1;1-x,-k_{\bot},\lambda_2 \rangle \, ,
\end{equation}
and for baryon states (with the three quark Fock space
sector),
\begin{eqnarray}
  |P^+,P_{\bot},SS_3\rangle &=& \sum_{\lambda_i} \int \prod_{i}^2
  \frac{\,dx_id^2k_{i\bot}\,}{16\pi^3}\Psi_b^{SS_3}(x_i,k_{i\bot},
  \lambda_i) \nonumber \\
 & & \times |x_1,k_{1\bot},\lambda_1;x_2,k_{2\bot},\lambda_2;
  1-x_1-x_2,-(k_{1\bot}+k_{2\bot}),\lambda_3 \rangle \, ,
\end{eqnarray}
where $\Psi^{SS_3}$ is the amplitude of the corresponding $q\bar{q}$ or
three quark sector (the wave function of the quark model):
\begin{equation}
  \Psi^{SS_3}={\cal F}\, \Xi^{SS_3}(k_{i\bot},\lambda_i)
  \Phi (x_i,k_{i\bot}) \, ,
\end{equation}
with ${\cal F}$ the flavor part of the wave function which is
the same as in the constituent quark model, and $\Xi$ and $\Phi$
are the spin and space parts that depend on the dynamics.
By ignoring the dynamic dependence of the spin configuration
and by using the Melosh transformation  [\cite{melosh}], 
\begin{equation}
  R_M(k_{i\bot},m_i) = \frac{\,m_i+x_iM_0-i\vec{\sigma}\cdot (\vec{n}
	\times \vec{k}_{i\bot})\,} {\sqrt{(m_i+x_iM_0)^2+k_{i\bot}^2\,}} \, ,
\end{equation}
where $\vec{n}=(0,0,1)$, $\vec{\sigma}$ is the Pauli spin matrix,
$m_i$ the $i$-th constituent quark mass, and $M_0$ satisfies
\begin{equation}
  M_0^2 = \sum_i \frac{\,k_{i\bot}^2+m_i^2\,}{x_i} \, ,
\end{equation}
the light-front spin wave function can be given by
\begin{eqnarray}
  \Xi_m^{SS_3}(k_{\bot},\lambda_1, \lambda_2) 
	&=& \sum_{s_1,s_2}\langle \lambda_1|R_M^{\dagger}
  (k_{\bot},m_1)|s_1\rangle \langle \lambda_2|R_M^{\dagger}(-k_{\bot},m_2)
  |s_2 \rangle \nonumber \\
	& & ~~~~~~~~~~~~~~~~~~~~~ \langle 1/2 s_1, 1/2 s_2 | SS_3 \rangle \, ,
\end{eqnarray}
for mesons; for baryons the spin part is rather complicated for a
detailed construction, see for example Ref. [\cite{Schlumpf}]. The 
momentum part of the wave function may be written as
\begin{equation}
  \Phi_m(x_i,k_{i\bot}) = {\cal N}_m \sqrt{dk_z\over dx} \exp 
	(-\vec{k}^2/2\omega^2_m) \,  
\end{equation} 
for meson, where $\vec{k}=(k_\bot, k_z), k_z= \Big(x-{1\over 2}\Big)M_0 - 
{m_1^2-m_2^2\over 2M_0}$; and for baryons 
\begin{equation}
  \Phi_b(x_i, k_{i\bot}) = {\cal N}_b \frac{1}{\,(1+M_0^2/
	\omega_b^2)^{3.5}\,} \, ,
\end{equation}
where ${\cal N}$ is a normalization constant and $\omega$ is a 
parameter fixed by the data. Other phenomenological light-front 
wave functions have also been used.

These phenomenological light-front wave functions have been widely 
used to calculate hadronic form factors and coupling constants 
 [\cite{plfwf}]; the results look pretty good for a very broad range 
of momentum transfer, and should provide a much better description 
than the nonrelativistic constituent quark model and other 
phenomenological descriptions.

Nevertheless, all these are just some phenomenological examinations of
light-front hadronic wave functions.  The true strong interaction
description of hadronic structure is the solution of the bound state
equation, Eq.(\ref{bse}) or approximately Eq.(\ref{lfbse}), from QCD. 
This is the main 
task of the recently development of QCD formulated on the light-front. 
In the remaining lectures, I will discuss the QCD formulation
on the light-front and then explore its application to heavy 
quark systems, based heavily on the works which have been done 
with my collaborators in the last few years
 [\cite{zhang93,zhang93a,zhang93b,hari2,Wilson94,zhang95,zhang96}].

\section{Canonical Light-Front QCD}

\subsection{Introduction}
Light-front QCD that I am going to discuss is the theory of QCD 
formulated on a light-front surface with the light-front gauge $A_a^+=0$. 
Before start the discussion on   LFQCD, I would like to
make a few remarks: First of all, I would like to claim that any 
problem of QCD that can be solved in the instant formulation should 
be undoubtedly solved on the light-front. This is not surprise at all!
However, the importance of   LFQCD is that we hope to solve the 
subtle problems in QCD that have not been solved in the instant form,
such as color confinement and dynamical chiral symmetry breaking
problems. To reach this goal, one may need to have some relatively
complete knowledge on the canonical formulation of   LFQCD 
and from which to
find the key problem associated with these subtleties. Hence,
in this lecture, I will introduce the canonical form of light-front
QCD, then discuss the origin of the light-front gauge singularity
and the light-front two-component formulation of QCD which has some very
special structure for field theory that are only manifested on the 
light-front.

The QCD Lagrangian is defined by
\begin{equation}
  {\cal L} = -  \frac{1}{\,2\,}{\rm Tr}(F^{\mu \nu}F_{\mu \nu})
  + \overline{\psi} (i\gamma_{\mu}D^{\mu}-m)\psi \, ,
\end{equation}
where $F^{\mu \nu}=\partial^{\mu}A^{\nu}-\partial^{\nu}A^{\mu}-
ig[A^{\mu},A^{\nu}]$, $A^{\mu}=\sum_a A_a^{\mu}T^a$ is a
$3\times 3$ gluon field color matrix and the $T^a$ are the generators
of the $SU(3)$ color group: $[T^a,T^b]=if^{abc}T^c$ and ${\rm Tr}
(T^aT^b)=\frac{1}{2}\delta_{ab}$. The field variable $\psi$
describes quarks with three colors and $N_f$ flavors,
$D^{\mu}=\partial^{\mu}-igA^{\mu}$ is the
symmetric covariant derivative, and $m$ is an $N_f\times N_f$
diagonal quark mass matrix.  The Lagrange equations of motion are
well-known:
\begin{eqnarray}
 &&  (i\gamma_{\mu}\partial^{\mu}-m+g\gamma_{\mu}A^{\mu})
  \psi = 0 \, , \label{dirace} \\
 &&  \partial_{\mu}F_a^{\mu \nu} +gf^{abc}A_{b\mu}F_c^{\mu \nu}
  + g\overline{\psi}\gamma^{\nu}T^a \psi =0 \, , \label{gfe}
\end{eqnarray}

The following discussion and also that of the next lecture are 
mainly based on the work in collaboration with A. Harindranath 
 [\cite{zhang93a,zhang93b,hari2}].

\subsection{Light-front (phase space) Quantization}
To formulate the QCD on the light-front, the following light-front 
notations will be adopted: The space-time coordinate is denoted by
$x^{\mu} = (x^+,x^-,x_{\bot})$, where $x^+ = x^0 + x^3$ is the 
light-front time-like component, $x^- = x^0 - x^3$ and $x_{\bot}^i~ 
(i=1,2)$ are respectively the light-front longitudinal and 
transverse components. The light-front derivatives are given by
$\partial^+=2{\partial\over\partial x^-}$, $\partial^-=2{\partial
\over\partial x^+}$, and $\partial_{\bot}^i={\partial\over\partial 
x^i}$. The product of two four-vectors is written as $a \cdot b = 
{1 \over 2}( a^+b^- + a^- b^+) - a_{\bot} \cdot b_{\bot}$.

The canonical theory of QCD on the light-front is constructed with
the choice of the light-front gauge $A_a^+ = 0$.  The first question
you may ask is why we choose the light-front gauge. The answer is 
as follows:

On the light-front, the quark (more generally the fermion) fields can 
be decoupled into $\psi(x) = \psi_+(x) + \psi_-(x) $ with $\psi_\pm(x)
= {1\over 2}\gamma^0 \gamma^\pm \psi(x)$. Then the Dirac equation
(\ref{dirace}) can be separated into:
\begin{eqnarray}
 &&  \Big(i\partial^- + gT^a A_a^-\Big) \psi_+ 
	= \Big(i\alpha_\bot \cdot D_\bot + \beta m\Big) \psi_- \, ,\\
 &&  \Big(i\partial^+ + gT^a A_a^-\Big) \psi_- 
	= \Big(i\alpha_\bot \cdot D_\bot + \beta m\Big) \psi_+ \, ,
\end{eqnarray}
where $\alpha_\bot = \gamma^0 \gamma_\bot, \beta = \gamma^0$.
It shows that the component $\psi_-$ is a constraint field variable,
which can be solved nonperturbatively from the about equation ONLY if
we take $A_a^+=0$!
\begin{equation}
  \psi_- = {1\over i \partial^+} \Big(i\alpha_\bot \cdot D_\bot + \beta 
	m\Big) \psi_+ \, . 
\end{equation}

Secondly, due to gauge symmetry among the four components of the
vector gauge filed, only two of them are the physically independent
variables. By taking $A_a^+=0$, the equation of motions (\ref{gfe}) 
for $\nu=+$ is simply reduced to
\begin{equation}
	{1\over 2} \Big(\partial^+\Big)^2 A_a^- = \partial^+ 
		\partial^i A_a^i + g \rho_a ,
\end{equation}
where $\rho_a = f^{abc} A_b^i \partial^+ A_c^i + 2 \psi_+^\dagger
T^a \psi_+$ is the light-front color charge density. This is indeed
the light-front Gauss Law which can be used to determines $A_a^-$ 
in terms of $\psi_+$ and $A_a^i$:
\begin{equation}
	A_a^- = 2 \Bigg\{\Big({1\over \partial^+}\Big) (\partial^i
		A_a^i) + \Big({1\over \partial^+}\Big)^2 \rho_a\Bigg\},
\end{equation}  
where the operator $\frac{1}{\partial^+}$ will be defined later.
It shows that with the light-front gauge, we can explicitly 
eliminate all the unphysical gauge degrees of freedom. 

Now, we can write a simple close form for the   LFQCD 
Lagrangian in terms of the pure physical degrees of freedom,
$\psi_+$ and $A_a^i~(i=1,2)$, with the choice of the light-front 
gauge:
\begin{equation}
	{\cal L}_{QCD} = {1\over 2} (\partial^+ A_a^i) (\partial^-
		A_a^i) + i\psi_+^\dagger \partial^- \psi_+
 		- {\cal H},
\end{equation}
where ${\cal H}$ is the   LFQCD Hamiltonian density: 	
\begin{eqnarray}
  {\cal H} = && \frac{1}{\,2\,}(\partial^i A_a^j)^2+gf^{abc}A_a^iA_b^j
	\partial^iA_c^j +  \frac{\,g^2\,}{4}f^{abc}f^{ade}A_b^i
	A_c^jA_d^iA_e^j  \nonumber \\ 
  && + \biggl [ \psi_+^{\dagger} \bigl \{ \alpha_{\bot}\cdot (i
	\partial_{\bot}+gA_{\bot})+ \beta m\bigr \} \biggl ( \frac{1}{\,
	i\partial^+\,}\biggr ) \bigl \{ \alpha_{\bot}\cdot (i
	\partial_{\bot}+g A_{\bot})+ \beta m\bigr \} \psi_+ \biggr ] 
		\nonumber \\
 && + g\partial^iA_a^i \biggl (  \frac{1}{\,\partial^+\,}
  \biggr ) \rho_a +  \frac{\,g^2\,}{2}\biggl (  \frac{1}{\,\partial^+\,}
  \biggr ) \rho_a \biggl (  \frac{1}{\,\partial^+\,}\biggr ) \rho_a .
\end{eqnarray}

Next we discuss the light-front quantization. A self-consistent 
canonical quantization requires that the resulting
Hamiltonian must generate the correct equations of motion for the
physical degrees of freedom $(A_a^i, \psi_+, \psi^{\dagger}_+)$.
To reproduce the Lagrangian equations of motion, we need to find
consistent commutators for physical field variables. In the
light-front gauge, the   LFQCD phase space is spanned 
by the field variables, $A_a^i, \psi_+, \psi_+^{\dagger} $ and 
their canonical momenta, ${\cal E}_a^i=\frac{1}{2}\partial^+ A_a^i$,
$\pi_{\psi_+} =\frac{i}{2}\psi_+^{\dagger}$, $\pi_{\psi^{\dagger}_+}
=-\frac{i}{2}\psi_+$. The phase space structure which determines 
the Poisson brackets of its variables can be found by the Lagrangian 
one-form ${\cal L}dx^+$ (apart from a total light-front time derivative),
\begin{eqnarray}
  {\cal L}dx^+ &=&  \frac{1}{\,2\,}2({\cal E}_a^i dA_a^i + \pi_{\psi_+}
  d\psi_+ + d\psi_+^{\dagger} \pi_{\psi_+^{\dagger}} \nonumber \\
	& & ~~~~~~~~~~~ - A_a^i
  d{\cal E}_a^i - d\pi_{\psi_+}\psi_+ - \psi_+^{\dagger}
  d\pi_{\psi_+^{\dagger}}) - {\cal H} dx^+  \nonumber \\ 
  &=&  \frac{1}{\,2\,} q^{\alpha} \Gamma_{\alpha \beta}
  dq^{\beta} - {\cal H} dx^+ \, ,
\end{eqnarray}
where the first term on the right-hand side is called the canonical
one-form of the phase space (note that quark fields are anticommuting
$c$-numbers (Grassmann variables)).  Correspondingly, the canonical 
equal-$x^+$ commutation relations are then given by:
\begin{equation}
  [q^{\beta}(x), q^{\alpha}(y)]_{x^+=y^+} = i \Gamma_{\alpha
  \beta}^{-1} \, .
\end{equation}
Explicitly, we have
\begin{eqnarray}
&&  \bigl \{ \psi_+(x),\psi_+^{\dagger}(y)\bigr \}_{x^+=y^+}
  = i\Lambda_+ \delta^3(x-y) \, , \\
&&   \bigl [ A_a^i(x),A_b^j(y)\bigr ]_{x^+=y^+}
  = i\delta_{ab}\delta^{ij}\, \Bigg({1\over \partial_y^+}\Bigg)
	\delta^3 (x - y) \, .
\end{eqnarray}
From these commutation relations it is straightforward to verify that
the Hamiltonian equations of motion are consistent with Eqs.(\ref{dirace})
and (\ref{gfe}). As we see in the above light-front quantization 
of QCD one does not need to introduce the ghost field. However, this 
canonical formulation does not completely define theory for practical 
computations due to existence of gauge singularity.  

\subsection{Light-Front Gauge Singularity}

The gauge singularity is perhaps the most difficult problem in 
non-abelian gauge theory that has not been completely solved since 
it was developed. In   LFQCD, it arises from the elimination 
of the unphysical gauge degrees of freedom. To eliminate the 
unphysical degrees of freedom on the light-front, we need to solve
the constraint equations which depend on the definition of the operator 
$1/\partial^+$. In our formulation, we define this operator by 
\begin{equation}
  \biggl (  \frac{1}{\,\partial^+\,}\biggr ) f(x^-,x^+,x_{\bot})
  =  \frac{1}{\,4\,}  \int_{-\infty}^{\infty} dx_1^-
  \varepsilon (x^- - x_1^-) f(x_1^-,x^+,x_{\bot}) \, , \label{1p+}
\end{equation}
where $\varepsilon (x) = -1, 0, 1$ for $x <0, =0, >0$. 

In perturbation theory, the gauge singularity manifests itself clearly 
in momentum space. The momentum representation of Eq.(\ref{1p+}) is
\begin{equation}
\begin{array}{rl}
  \!\!\! \biggl (  \frac{1}{\,\partial^+\,}\biggr )^nf(x^-) = \!\!\!\! & \biggl (
   \frac{1}{\,4\,}\biggr )^n \int_{-\infty}^{\infty}dx_1^-\cdots dx_n^-
  \epsilon (x^- - x_1^-)\cdots \epsilon (x_{n-1}^-
  -x_n^-)f(x_n^-) \\ 
  & \longrightarrow \biggl [  \frac{1}{\,2\,}\biggl (
   \frac{1}{\,k^++i\epsilon\,}+ \frac{1}{\,k^+-i\epsilon\,}\biggr )
  \biggr ]^nf(k^+)= \frac{1}{\,[k^+]^n\,}f(k^+) \, . \label{1p+1}
\end{array}
\end{equation}
As we see the $k^+=0$ modes are removed with this definition. In other
words, the singularity of $\frac{1}{k^+}$ is regularized. However,
with such an infrared regularization, many infrared divergences from 
the small longitudinal momentum, surrounding the $k^+=0$ region,
will occurs in the perturbative calculation. We will discuss these 
divergences in the next lecture.

On the other hand, with the definition of Eq.(\ref{1p+}), we have
\begin{equation}
  \bigl [ A_a^i(x),A_b^j(y)\bigr ]_{x^+=y^+} = -i {1\over 4} 
	\delta_{ab}\delta^{ij}\, \epsilon(x^- - y^-) \delta^2
  	(x_{\bot} - y_{\bot}) \, .
\end{equation}
This leads to the fact that $A_a^i$ satisfies an antisymmetric 
boundary condition:
\begin{equation}
	A_a^i(x^-=-\infty) = -A_a^i(x^-=+\infty).
\end{equation}
It also shows that the zero-mode (the longitudinal momentum is zero)
in $A_a^i$ is removed. Meanwhile, quarks in QCD should always
be massive, namely their longitudinal momentum is not really zero. 
Thus, with the definition of Eq.(\ref{1p+}), the theory of light-front 
QCD does not contain zero-modes. Therefore the   LFQCD vacuum 
in this formulation is always trivial! Now you may ask where is the 
nontrivial properties of QCD with such a trivial vacuum in your
formulation?

Apparently, after solving $A_a^-$ component from the light-front Gauss 
law in the $A_a^+=0$ gauge, the gauge freedom should be completely 
fixed. However, a careful check shows that there is still a residual
gauge transformation in the above formulation. It is given by
\begin{equation}
	U_r = \exp \Bigg\{ - {i\over g} \int d^2 x_\bot 
		\theta^a(x_\bot) R_a(x_\bot) \Bigg\},
\end{equation}
which is associated with the field $A_a^-$ at the longitudinal infinity:
\begin{equation}
	R_a = {1\over 2}\partial^+ A_a^-|_{x^-=\infty} = 
		{1\over 2}\int_{-\infty}^{\infty} dx^-\Big[2\partial^+
		\partial^i A_a^i(x) + g\rho_a(x) \Big] .
\end{equation}
 
This gauge freedom can be further fixed for physical states.
This is because the operator $R_a = E_a^-|_{x^-=\infty}$ which is
the longitudinal component of color electric field strength at 
longitudinal infinity. For physical states, finite energy density 
requires that the color electric field strength must vanish at
the longitudinal boundary: $E_a^- |_{x^- = \pm \infty} = 0$. This 
condition canonically removes the residual gauge freedom and 
leads to a constraint on the $A_a^i$ at longitudinal infinity:
\begin{equation}
  \partial^i A_a^i |_{x^- = \pm \infty} = \mp  \frac{\,g\,}{2} 
	 \int_{-\infty}^{\infty} dx^- \rho_a(x^-,x) \, .
\end{equation}
The nontrivial properties of QCD in our formulation are indeed
hidden in this condition. The main effect of this equation should 
be only manifested in nonperturbative dynamics, i.e., in physical 
bound states. An explicit nontrivial effect can be seen form the 
axial anomaly of QCD, for example. Consider the axial current 
equation (for zero quark mass) 
\begin{equation}
  \partial_{\mu} j_5^{\mu} = N_f  \frac{g^2}{\,8\pi^2\,}\,
  \mbox{Tr}\,(F_{\mu \nu} \widetilde{F}^{\mu \nu})\, ,
\end{equation}
where the axial current is $j_5^{\mu}=\bar{\psi}\gamma^{\mu}\gamma_5 
\psi$, and the dual field strength is $\widetilde{F}^{\mu\nu}=\frac{1}
{2}\epsilon^{\mu\nu\sigma\rho}F_{\sigma\rho}$. The winding number in 
  LFQCD is defined as the net charge between
$x^+=-\infty$ and $x^+=\infty$,
\begin{equation}
  \Delta Q_5 = N_f  \frac{g^2}{\,8\pi^2\,} \int_M d^4x\,\mbox{Tr}\,
  (F_{\mu \nu}\widetilde{F}^{\mu \nu})\, .
\end{equation}
The integration on the r.h.s. of the above equation is defined in
Minkowski space ($M$) and can be replaced by a surface integral.  
It has been found  [\cite{zhang93}] that
\begin{equation}
  \Delta Q_5 = -N_f \frac{\,g^2\,}{\,\pi^2\,} \int dx^+ d^2x_{\bot}\,
  	\mbox{Tr}\,\bigl ( A^-[A^1,A^2]\bigr ) \Bigl 
	|_{x^-=-\infty}^{x^-=\infty} \, , 
\end{equation}
namely, a non-vanishing $\Delta Q_5$ is generated from the asymptotic 
fields of $A_a^i$ and their antisymmetric boundary conditions at 
longitudinal infinity.

From the above canonical analysis, we can see that nontrivial 
features in   LFQCD are induced by the gauge singularity 
and are manifested at the longitudinal infinity on the light-front.
They are also associated with the light-front longitudinal
infrared divergence in momentum space when the zero-modes are 
removed in our canonical quantization. This analysis gives us
some hint where we should look for the problems in the study 
of nonperturbative QCD with a trivial vacuum on the light-front.

\subsection{Two-Component Formulation}
 
When QCD is formulated on the light-front, the theory can be expressed 
in terms of a pure two-component form.  This is another useful feature 
of   LFQCD. After the elimination of the unphysical gauge
degrees of freedom, the QCD gauge field has already been reduced
to the two transverse components, $A_a^1$ and $A_a^2$. While, as
we will see soon that the quark field can also be written in terms
of a two-component field (rather than the four-component field
in instant form)  [\cite{zhang93b}].

To do so, we should introduce the following $\gamma$ matrix
representation:
\begin{equation}
	\gamma^0=\left[\begin{array}{cc} 0 & -i \\ i & 0 \end{array}
		\right]~,~~~
	\gamma^3=\left[\begin{array}{cc} 0 & i \\ i & 0 \end{array}
		\right]~,~~~
	\gamma^i=\left[\begin{array}{cc} -i\epsilon^{ij}\sigma_j & 0 
		\\ 0 & i\epsilon^{ij}\sigma_j \end{array}\right] .
\end{equation}
Then, one can find that the light-front project operators become
\begin{equation}
	\Gamma^+={1\over 2} \gamma^0 \gamma^+ = \left[\begin{array}{cc} 
		1 & 0 \\ 0 & 0 \end{array} \right]~,~~~
	\Gamma^-={1\over 2} \gamma^0 \gamma^- = \left[\begin{array}{cc} 
		0 & 0 \\ 0 & 1 \end{array} \right]~.
\end{equation}
and the light-front quark field have the two-component form:
\begin{equation}
	\psi= \left[\begin{array}{c} \varphi \\ \nu \end{array} \right]~,~~~
	\psi_+= \left[\begin{array}{c} \varphi \\ 0 \end{array} \right]~,~~~
	\psi_-= \left[\begin{array}{c} 0 \\ \nu \end{array} \right]
		= \left[\begin{array}{c} 0 \\ {1\over \partial^+} 
		\Big((D_\bot \times \sigma_\bot)^3 + m\Big)\varphi 
		\end{array} \right]~,
\end{equation}
where $\varphi(x)$ is a two-component spinor field. In the above 
expressions, $\sigma_i$ are the Pauli matrix. Thus, the 
relativistic fermion particles can be described as a nonrelativistic
spin ${1\over2}$ particle on the light-front. The canonical
commutation relation is also reduced to:
\begin{equation}
	\Big\{ \varphi(x)~,~\varphi^\dagger(y) \Big\}_{x^+=y^+} = 
		\delta^3 (x-y) .
\end{equation}

With the above formulation, the   LFQCD Hamiltonian can be 
rewritten as
\begin{equation}
  H =  \int dx^- d^2 x_{\bot} ({\cal H}_0 + {\cal H}_{int} )
  = H_0 + H_I \, ,
\end{equation}
where
\begin{eqnarray}
&&  {\cal H}_0 =  \frac{1}{\,2\,}(\partial^iA_a^j)(\partial^i A_a^j)
  +\varphi^{\dagger}\biggl (  \frac{\,-\nabla^2 + m^2\,}{i\partial^+}
  \biggr ) \varphi \, , \\
&&  {\cal H}_{int} = {\cal H}_{qqg} + {\cal H}_{ggg} +
  {\cal H}_{qqgg} + {\cal H}_{qqqq} + {\cal H}_{gggg} \, ,
\end{eqnarray}
and
\begin{eqnarray}
  {\cal H}_{qqg} = & & g\varphi^{\dagger}\biggl \{ - 2\biggl (
   \frac{1}{\,\partial^+\,}\biggr ) (\partial \cdot
  A_{\bot})+\sigma \cdot A_{\bot}\biggl (
   \frac{1}{\,\partial^+\,}\biggr ) (\sigma \cdot \partial_{\bot}
  +m)  \nonumber \\
  & & ~~~~~~~~ + \biggl (  \frac{1}{\,\partial^+\,}\biggr )
  (\sigma \cdot \partial_{\bot}-m)\sigma \cdot A_{\bot}
  \biggr \} \varphi \, , \\
  {\cal H}_{ggg} = && g f^{abc} \biggl \{ \partial^i A_a^j A_b^i A_c^j
  + (\partial^i A_a^i) \biggl (  \frac{1}{\,\partial^+\,} \biggr )
  (A_b^j \partial^+ A_c^j) \biggr \} \, , \\
  {\cal H}_{qqgg} = & & g^2 \biggl \{ \varphi^{\dagger} \sigma \cdot
  A_{\bot} \biggl (  \frac{1}{\,i\partial^+\,}\biggr ) \sigma \cdot
  A_{\bot}\varphi \nonumber \\
  & & ~~~~~~~~ +2\biggl (  \frac{1}{\,\partial^+\,}
  \biggr ) (f^{abc}A_b^i\partial^+A_c^i)\biggl (  \frac{1}{\,\partial^+\,}
  \biggr ) (\varphi^{\dagger}T^a\varphi ) \biggr \} \nonumber \\
  = && {\cal H}_{qqgg1} + {\cal H}_{qqgg2} \, , \\
  {\cal H}_{qqqq} = && 2g^2 \biggl \{ \biggl (  \frac{1}{\,\partial^+\,}
  \biggr )(\varphi^{\dagger}T^a\varphi ) \biggl (  \frac{1}{\,\partial^+\,}
  \biggr ) (\varphi^{\dagger}T^a\varphi ) \biggr \} \, , \\
  {\cal H}_{gggg} = &&  \frac{\,g^2\,}{4}f^{abc}f^{ade}
  \biggl \{ A_b^iA_c^jA_d^iA_e^j+2\biggl (  \frac{1}{\,\partial^+\,}
  \biggr )(A_b^i\partial^+A_c^i)\biggl (  \frac{1}{\,\partial^+\,}
  \biggr ) (A_d^j\partial^+A_e^j)\biggr \} \nonumber \\
  = &&  {\cal H}_{gggg1} + {\cal H}_{gggg2} \, .
\end{eqnarray}

The above two-component formulation simplify the relativistic field
theory structure, especially in the study of the relativistic bound 
state problems.

\section{LF Time-Ordered Perturbation Theory for QCD}

\subsection{About Light-Front Perturbative QCD}

Time-ordered perturbation theory, especially the light-front
time-ordered perturbation theory, provides a natural perturbative
description for parton phenomena  [\cite{Yan,Miller}]. The current
attempts of solving nonperturbative QCD dynamics on light-front 
is also based on the analysis of time-ordered approach in 
Hamiltonian formulation. However,
the light-front gauge singularity discussed above will lead to
severe infrared divergences in such perturbation theory, although
Eq.(\ref{1p+1}) provides a well-defined regulator (a generalized 
principal value prescription) for the small $k^+$ momentum.
  
In covariant perturbation theory, the use of the principal value
prescription still leads to the so-called ``spurious'' poles in the 
light-front 
Feynman integrals, which prohibit any continuation to Euclidean 
space (Wick rotation) and hence the use of standard power counting 
arguments for Feynman loop integrals.  This causes difficulties in 
addressing renormalization of QCD in covariant perturbation theory 
with the light-front gauge. In the last decade there are many 
investigations attempting to solve this problem.  One excellent 
solution is given by Mandelstam and Leibbrandt, i.e., the 
Mandelstam-Leibbrandt (ML) prescription  [\cite{MLd}], which allows 
continuation to Euclidean space and hence power counting. It has also 
been shown that, with the ML prescription, the {\em multiplicative}
renormalization in the two-component   LFQCD Feynman 
formulation is restored  [\cite{Lee86}].

Unfortunately, the ML prescription cannot be applied to equal-$x^+$ 
quantization because the ML prescription is defined by a boundary 
condition which depends on $x^+$ itself and is not allowed in 
equal-$x^+$ canonical theory. Yet, as we pointed out recently 
 [\cite{Wilson94}], light-front power counting differs completely 
from the power counting in equal-time quantization that noncanonical 
counterterms are allowed in light-front field theory.  In other words,
multiplicative renormalization is not required in   LFQCD.  
Furthermore, the current attempts to understand nonperturbative 
QCD in light-front coordinates is based on the $x^+$-ordered 
diagrams in which no Feynman integral is involved. Thus the power 
counting criterion for Feynman loop integrals is no longer available in
  LFQCD Hamiltonian calculations. In $x^+$-ordered perturbation
theory with the principle value prescription,   LFQCD contains
severe linear and logarithmic infrared divergences. Here I will give 
some results from the $x^+$-ordered perturbative loop calculations and 
renormalization of   LFQCD Hamiltonian theory up to one-loop 
 [\cite{zhang93b,hari2}], where the infrared divergences are 
systematically analyzed. 
Since light-front power counting allows noncanonical counterterms, a 
complete understanding of renormalized   LFQCD may not be 
worked out within perturbation theory; new renormalization and 
regularization approaches need to be developed, as we will see
later.

\subsection{LF $x^+$-Ordered Perturbation Theory}

The $x^+$-ordered perturbation theory can be obtained 
from the familiar perturbation expansion in quantum mechanics.
The perturbation expansion of a bound state is given by
(in the Rayleigh-Schr\"{o}dinger perturbation theory):
\begin{equation}
  |\Psi \rangle =  \sum_{n=0}^{\infty}\biggl (  \frac{Q}{\,E_0-H_0\,}
  (H'_I)\biggr )^n|\Phi \rangle \, ,
\end{equation}
where $|\Phi \rangle$ is a unperturbative state, $Q$ and $H'_I$
are defined by:
\begin{equation}
  Q = |\Phi \rangle \langle \Psi |\, , \quad \quad H'_I=H_I-\Delta E\, , \quad \quad
  \Delta E=\langle \Phi |H_I|\Psi \rangle \, .
\end{equation}
With this perturbative expansion formula, the mass, the wave functions,
and the coupling constants renormalizations can be expressed
as follows.  For the convenience of practical calculations,
we consider the expressions in momentum space.

\hspace{0.3cm} i). {\em Wavefunction renormalization}: In momentum space, the
perturbative expansion of a state is given by
\begin{eqnarray}
  |\Psi \rangle &= & \Biggl \{ |\Phi \rangle +{ \sum_{n_1}}' \frac{\,
	|n_1\rangle \langle n_1|H'_I|\Phi \rangle\,}{\,p^--p_{n_1}^- \,} 
	\nonumber \\  
  & & ~~~~~ + { \sum_{n_1n_2}}'  \frac{|n_1\rangle \langle n_1|H'_I|n_2
	\rangle \langle n_2|H'_I|\Phi \rangle\,} {\,(p^--p_{n_1}^-)(p^-
	-p_{n_2}^- )\,}+\cdots \Biggr \} \, , \label{wfr}
\end{eqnarray}
which has not been normalized, where $| n_1 \rangle, | n_2 \rangle,
\cdots $ are properly symmetrized (antisymmetrized) states with
respect to identical bosons (fermions) in the states and $\sum'$ in
Eq.(\ref{wfr}) sums over all intermediate states except the initial
state $| \Phi \rangle$. The normalized wave function is defined by
$| \Psi' \rangle = \sqrt{Z_{\Phi}} | \Psi \rangle $,
where the factor $Z_{\Phi}$ is the wavefunction renormalization
constant:
\begin{equation}
  Z_{\Phi}^{-1} = \langle \Psi | \Psi \rangle
  = 1 +  \sum_{n_1}{'} \frac{|\langle n_1|H'_I|\Phi \rangle|^2\,}
  {\,(p^--p_{n_1}^- )^2\,}+\cdots \, .
\end{equation}

\hspace{0.3cm} ii). {\em Mass renormalization}.
The mass correction can then be computed from the ``energy-level'' shift,
i.e., the correction to the energy of an on-mass-shell particle.  It
is obvious that the perturbative correction to the light-front energy
($p^-$) is given by
\begin{eqnarray}
  \delta p^- & = & \langle \Phi |(H-H_0)|\Psi \rangle
  = \langle \Phi |H_I|\Psi \rangle \nonumber \\ 
  & = & \langle \Phi |H_I|\Phi \rangle
  +{ \sum_{n_1}}' \frac{\,|\langle n_1|H_I|\Phi \rangle 
	|^2\,}{\,p^--p_{n_1}^- \,} +\cdots \, .
\end{eqnarray}
Using the mass-shell equation $m^2=p^+p^--{\bf p}_{\bot}^2$, and recalling
that $p^+$ and ${\bf p}_{\bot}$ are the conserved light-front kinematical
momenta, we obtain the mass renormalization in the old-fashioned
perturbative light-front field theory:
\begin{equation}
  \delta m^2=p^+ \delta p^- =  p^+ \langle \Phi |H_I
  |\Phi \rangle + p^+ { \sum_{n_1}}'
   \frac{\,|\langle n_1|H_I|\Phi \rangle |^2\,}{\,p^--p_{n_1}^- \,}
  +\cdots \, .
\end{equation}

\hspace{0.3cm} iii). {\em Coupling constant renormalization}. The coupling
constant renormalization is obtained by the perturbative calculation
of various matrix elements of the vertices in $H_I$.  Consider
a vertex $H_I^i$ that is proportional to the coupling constant
$g$, we have
\begin{eqnarray}
  \langle \Psi'_f|H_I^i|\Psi'_i \rangle & \equiv & Z_g
  \sqrt{Z_iZ_f\,}\langle \Psi_f|H_I^i|\Psi_i \rangle  \nonumber \\ 
  & = & \langle \Phi_f|H_I^i|\Phi_i \rangle
  + { \sum_{n_1}}' \frac{\,\langle \Phi_f|H'_I|n_1\rangle \langle 
	n_1|H_I^i|\Psi_i\rangle\,}{p_f^--p_{n_1}^- }  \nonumber\\  
  & & + { \sum_{n_1}}' \frac{\,\langle \Phi_f|H_I^i|n_1\rangle 
	\langle n_1|H'_I|\Psi_i\rangle\,}{p_i^--p_{n_1}^-}  \nonumber\\  
  & & + { \sum_{n_1,n_2}}' \frac{\,\langle \Phi_f|H'_I|n_1\rangle 
	\langle n_1|H'_I|n_2\rangle \langle n_2|H_I^i]|\Phi_i \rangle\,}
  	{(p_f^--p_{n_1}^- )(p_f^--p_{n_2}^-+i\epsilon )}  \nonumber\\  
  & & + { \sum_{n_1,n_2}}' \frac{\,\langle \Phi_f|H'_I|n_1\rangle 
	\langle n_1|H_I^i|n_2\rangle \langle n_2|H'_I]|\Phi_i \rangle\,}
  	{(p_f^--p_{n_1}^- )(p_i^--p_{n_2}^- )}  \nonumber\\  
  & & + { \sum_{n_1,n_2}}' \frac{\,\langle \Phi_f|H_I^i|n_1\rangle 
	\langle n_1|H'_I|n_2\rangle \langle n_2|H'_I]|\Phi_i \rangle\,}
  {(p_i^--p_{n_1}^- )(p_i^--p_{n_2}^- )} +\cdots \, ,
\end{eqnarray}
where $Z_g$ is the multiplicative coupling constant renormalization,
and $Z_i$ and $Z_f$ are the wavefunction renormalization constants of
the initial and final states.

It is also convenient to express the above perturbation expansion 
in terms of the diagrammatic approach.  The rules for writing the
expression of perturbative expansions from diagrams for QCD are 
as follows:

\begin{enumerate}
\item [{$\bullet$}] Draw all topologically distinct $x^+$-ordered diagrams.

\item [{$\bullet$}] For each internal line, sum over helicity and 
integrate using $\int \frac{dk^+d^2k_{\bot}}{16\pi^3}\theta (k^+)$ for 
quarks and $\int \frac{dk^+d^2k_{\bot}}{16\pi^3 k^+}\theta (k^+)$ for
gluons.

\item [{$\bullet$}] For each vertex, include a factor of
$16\pi^3\delta^3(p_f-p_i)$ and a simple matrix element listed in Ref.
 [\cite{zhang93b}].

\item [{$\bullet$}] Include a factor $(p_i^--\sum_n p_n^-+i\epsilon )^{-1}$
[or $(p_f^--\sum_n p_n^-+i\epsilon )^{-1}]$ for each
intermediate state, where $\sum_n p_n^-$ sum over all on-mass-shell
intermediate particle energies.
\item [{$\bullet$}] Add a symmetry factor $S^{-1}$ for each gluon 
loop coming from the symmetrized boson states.
\end{enumerate}

\subsection{Perturbative Calculation of Light-Front QCD}
To illustrate the above computation scheme and to explore the severe 
light-front infrared divergences, let me list some calculations up to 
one-loop based on the $x^+$-ordered diagrammatical approach. Note 
that besides the infrared divergence, which is regularized by 
Eq.(\ref{1p+1}), there are also ultraviolet divergences for which 
we use a transverse cut-off: $\Lambda_{\bot}^2 \geq \kappa_{\bot}^2 
\geq \mu^2$. Here I have also introduced a mass scale $\mu$ for the 
minimum cut-off of the transverse momentum in order to avoid the 
several complicated pure infrared divergences and mass singularity 
from the massless gluon, $\mu$ should be much larger than all other 
masses in the theory, and is considered as a renormalization scale 
here.

\hspace{0.3cm} i). {\em Quark wavefunction and mass renormalization}. 
The one-loop light-front quark energy corrections (for the three 
diagrams in Fig. 1, respectively) are given by
\begin{eqnarray}
  \delta p_1^- &=& -  \frac{g^2}{\,8\pi^2\,}C_f\biggl \{ \frac{\,
  p^2-m^2\,}{[p^+]} \biggl ( 2\ln  \frac{\,p^+\,}{\epsilon}
  - \frac{\,3\,}{2}\biggr ) \ln {\Lambda_{\bot}^2 \over \mu^2} 
	 \nonumber \\
  & & ~~~ +  \frac{m^2}{\,[p^+]\,}\biggl ( -2 \ln {\Lambda_{\bot}^2 \over 
	\mu^2} \biggr ) + \frac{\,\Lambda_{\bot}^2 - \mu^2\,}{\,[p^+]\,}
	\biggl (  \frac{\,\pi p^+\,}{2\epsilon} -1+\ln  \frac{\,
	p^+\,}{\epsilon}\biggr ) \biggr \} \, , \\
  \delta p_2^- &=&  \frac{g^2}{\,8\pi^2\,}C_f  \frac{\,\Lambda_{\bot}^2
	-\mu^2 \,} {\,[p^+]\,}\ln  \frac{\,p^+\,}{\epsilon}\, , \nonumber \\
  \delta p_3^- &=&  \frac{g^2}{\,8\pi^2\,}C_f  \frac{\,\Lambda_{\bot}^2
	-\mu^2\,} {\,[p^+]\,}\biggl (  \frac{\,\pi p^+\,}{2\epsilon}-1 
	\biggr ) \, .
\end{eqnarray}
This shows that, in the one-loop quark energy correction, one-gluon 
exchange gives rise to both linear and logarithmic infrared
divergences.  The instantaneous fermion interaction contribution
(see $\delta p_2^-$ in Fig. 1b) contains only one logarithmic
divergence which cancels the logarithmic divergence in $\delta p_1^-$.
The instantaneous gluon interaction contribution ($\delta p_3^-$
of Fig. 1c) has a linear infrared divergence which precisely
cancels the same divergence in $\delta p_1^-$.  This cancellation 
of linear infrared divergences is based on the use of the 
regularization for $k^+ \rightarrow 0 $ in Eq. (\ref{1p+1})
 [\cite{zhang93b}].

\input psbox.tex
\begin{figure}[htbp] 
\begin{center}
\mbox{\psboxto(5.0in;1.2in){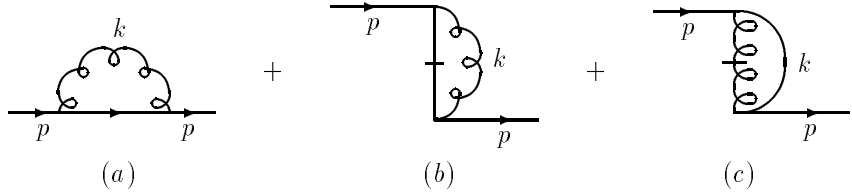}}
\end{center}
\caption{The $x^+$-ordered graphs for the one-loop correction to the
quark mass and wave function renormalization.}
\end{figure}

The quark mass correction (dropping the finite part) is then
given by
\begin{equation}
  \delta m^2 = p^+ \delta p^- |_{p^2=m^2} =  \frac{g^2}{\,4\pi^2\,} C_f
  m^2\ln  \frac{\,\Lambda_{\bot}^2\,}{\,\mu^2\,} \, ,
\end{equation}
which is longitudinally infrared divergence free; and the quark
wavefunction renormalization constant is
\begin{equation}
  Z_{2} = \!\!  1+ \frac{\,\partial \delta p^-\,}{\partial p^-}
  \biggr |_{p^2=m^2} = 1 +  \frac{g^2}{\,8\pi^2\,}C_f \biggl ( 
	 \frac{\,3\,}{2} -2\ln  \frac{\,p^+\,}{\epsilon}
	\biggr ) \ln  \frac{\,\Lambda_{\bot}^2\,}{\,\mu^2\,}\, .
\end{equation}
The wavefunction renormalization contains an additional
type of divergence, the mixing of infrared and ultraviolet
divergences, that does not occur in covariant calculations.
This is the `spurious' mixing associated with the gauge singularity.
It corresponds to the so-called light-front double pole problem 
in the Feynman theory with the use of the light-front gauge and
the principal value prescription that prohibits any continuation to 
Euclidean space and power counting in Feynman loop integrals.  
In the $x^+$-ordered Hamiltonian perturbation theory the power 
counting is different. The above argument$\,$ of$\,$ power$\,$ 
counting$\,$ for Feynman loop integrals may be irrelevant. 
Furthermore, since the second order correction to wavefunctions 
must be negative, the above result shows that it is the additional 
infrared divergence that gives a consistent answer for 
wavefunction renormalization.

\hspace{0.3cm} ii). {\em Gluon wave function and mass correction}. 
Similar calculation to the one-loop light-front gluon energy 
corrections leads to the solution:
\begin{eqnarray}
  \delta \mu_G^2 &=& -  \frac{g^2}{\,4\pi^2\,}\biggl \{ T_fN_f m^2\ln
   \frac{\,\Lambda_{\bot}^2\,}{\,\mu^2\,}(\Lambda_{\bot}^2-\mu^2)
  \biggl (  \frac{\,C_A\,}{2}-T_fN_f\biggr ) \biggl ( 1-\ln
   \frac{\,k_{\infty}^+\,}{\epsilon}\biggr ) \biggr \} \, , 
	\label{gmassc} \nonumber \\
  Z_3 &=& 1 +  \frac{g^2}{\,8\pi^2\,}\biggl \{ C_A \biggl ( 
   \frac{\,11\,}{6} -2\ln  \frac{\,q^+\,}{\epsilon}\biggr ) 
  - \frac{2}{\,3\,} T_fN_f\biggr \} \ln  \frac{\,\Lambda_{\bot}^2\,}
{\,\mu^2\,} \, . 
\end{eqnarray}

In the gluon sector, more severe divergences appear. It contains
the quadratic and logarithmic UV divergences, linear and 
logarithmic IR divergences, and an unusual large longitudinal
momentum logarithmic divergence.   Only the linear infrared 
divergences are cancelled with the principal value prescription.  
The gluon mass correction is not zero. The non-zero gluon mass 
correction of Eq.(\ref{gmassc}) is not surprising because it 
has the same 
divergence feature as the photon mass correction in light-front QED 
[Eq.(\ref{gmassc}) will be reduced to the photon mass correction when
we set $T_f=1$, $C_A=0$ and $N_f=1$].  In a covariant calculation,
the zero gluon mass correction is true only for dimensional
regularization which ``removes'' or drops the mass correction.
In the present calculation, maintaining zero gluon mass requires 
a mass counterterm, as is known in QED. The difference between 
QED and QCD is only manifest in the gauge boson wavefunction 
renormalization. For wavefunction renormalization, again there is an 
additional mixing of UV and IR divergences, which again provides the 
correct sign for the wavefunction renormalization constant.  

\hspace{0.3cm} iii). {\em Coupling constant renormalization}.  
For convenience, we set the external gluon momentum  
$q(q^+,q_{\bot}^i)=0$. The quark-gluon vertex is then reduced:
\begin{equation}
  {\cal V}_0 = 2gT_{\beta \alpha}^a \frac{p^i}{\,p^+\,}
  \delta_{\lambda_1\lambda_2}\varepsilon_{\sigma}^{i*} \, .
\end{equation}
In $x^+$-ordered perturbation theory, the one-loop vertex correction
is given by
\begin{equation}
  \delta {\cal V}_{0} = \{ V_1+V_2+V_{3}+V_{4}+V_{5}+V_{6} \}
  {\cal V}_{0} \, ,
\end{equation}
where $V_{n}$, $n=1-6$ are represented the contributions from 
different time-ordered diagrams (see Ref. [\cite{hari2}]):
\begin{eqnarray}
  V_1 &=&  \frac{g^2}{\,2\pi^2\,}\biggl (  \frac{\,3\,}{2}
  	-2\ln  \frac{\,p^+\,}{\epsilon}\biggr ) C_f\ln  \frac{\,
	\Lambda\,}{\,\mu\,} \, , \nonumber \\
  V_2 &=&  \frac{g^2}{\,8\pi^2\,}\biggl (  \frac{\,11\,}{3}C_A
  - \frac{\,4\,}{3}N_fT_f\biggr ) \, , \nonumber \\
  V_3 &=& -  \frac{g^2}{\,4\pi^2\,}\biggl (  \frac{\,3\,}{2}
  - 2\ln  \frac{\,p^+\,}{\epsilon}\biggr ) \biggl ( - \frac{1}{\,2\,}C_A
  +C_f\biggr ) \ln  \frac{\,\Lambda\,}{\,\mu\,} \, ,  \\
  V_4 &=& -  \frac{g^2}{\,8\pi^2\,}\biggl (  \frac{\,3\,}{2}
  - 2\ln  \frac{\,p^+\,}{\epsilon}\biggr ) C_A\ln  \frac{\,
	\Lambda\,}{\mu} \, , \nonumber \\
  V_5 &=& 0\, , \quad \quad V_{6}=0 \, .\nonumber
\end{eqnarray}

To evaluate the contributions to the coupling constant we
have to multiply $V_1$ and $V_2$ by $\frac{1}{2}$ in order to
 take into account the proper correction due to the
renormalization of initial and final states.
Thus adding the contributions together, we have,
\begin{equation}
\begin{array}{lll}
  \delta {\cal V}'_{0} & = & \biggl (  \frac{1}{\,2\,}V_1+ \frac{1}
	{\,2\,}V_2 +V_2+V_4+V_5+V_6\biggr ){\cal V}_{0} \\ 
  & = & {\cal V}_{0} \frac{g^2}{\,8\pi^2\,}\biggl (  \frac{\,11\,}{6} C_A
  -  \frac{2}{\,3\,}N_fT_f\biggr ) \ln  \frac{\,\Lambda\,}{\,\mu\,} \, 
	= \delta g {\cal V}_0.
\end{array}
\end{equation}
Note that all mixed divergences cancel now.
The correction to the coupling constant is given by
\begin{equation}
  g_R = g(1+\delta g) =g\biggl \{ 1+  \frac{g^2}{\,8\pi^2\,}
  \biggl (  \frac{\,11\,}{6}C_A -  \frac{2}{\,3\,} N_f
  T_f\biggr ) \ln  \frac{\,\Lambda\,}{\,\mu\,}\biggr \} \, .
\end{equation}
By redefining the bare coupling constant $g$ such that $g_R$ is finite.
Thus we have given all canonical renormalization quantities in QCD up to 
one-loop order based on the $x^+$-ordered perturbation theory.

From these results, the anomalous dimensions for quarks and gluons
and the $\beta$ function up to one-loop can be easily calculated.
The anomalous dimension of the quark field to order $g^2$ is
\begin{equation}
  \gamma_F \equiv  \frac{1}{\,2Z_2\,}  \frac{\partial Z_2}{\,
	\partial \ln \mu\,}
  =  \frac{g^2}{\,8\pi^2\,}C_f \biggl ( 2\ln  \frac{\,p^+\,}{\epsilon}
  -  \frac{\,3\,}{2} \biggr ) \, .
\end{equation}
The momentum-dependent term implies that the quark anomalous dimension
is gauge dependent.  The anomalous dimension for the gluon field is
\begin{equation}
  \gamma_G \equiv  \frac{1}{\,2Z_3\,} \frac{\partial Z_3}{\,\partial 
	\ln \mu\,} =  \frac{g^2}{\,8\pi^2\,}\biggl \{ C_f\biggl ( 2
  \ln  \frac{\,q^+\,}{\epsilon} -  \frac{\,11\,}{6} \biggr )
  + \frac{2}{\,3\,}T_fN_f\biggr \} \, , \label{adg}
\end{equation}
which is also gauge-dependent. In the case of $q^+=0$,
the gauge dependent term can be removed, and Eq.(\ref{adg}) is reduced
to Gross and Wilczek's result in their Feynman calculation with 
$A_a^+=0$ and $q^+=0$  [\cite{Gross74}].  The $\beta$ function is
\begin{equation}
  \beta (g) =  \frac{\partial g_R}{\,\partial \ln {\mu}\,}
  = -  \frac{g^3}{\,16 \pi^2\,}\biggl (  \frac{\,11\,}{3}C_A
  - \frac{4}{\,3\,}N_fT_f\biggr ) \, ,
\end{equation}
which is the well-known result to one loop order and is infrared
divergence free, as we expected.

From the above result, we see that there are severe light-front
divergences in   LFQCD.  Systematic control of these
divergences is required {\em a priori} before we perform any
practical numerical calculation in light-front coordinates for
QCD bound states.  From the basic one-loop calculations, one can
see that, in the $x^+$-ordered perturbation theory, light-front 
QCD involves various UV and IR divergences. Some of the 
divergences have not even been encountered in covariant and 
noncovariant Feynman calculations to the same order. Among 
various light-front divergences, there are two severe divergences
one has to deal with in the $x^+$-ordered theory for light-front 
QCD. The first is the mixing of UV and IR logarithmic divergences
in wavefunction renormalization.  The occurrence of the mixing 
divergences may not be a severe problem. The mixing divergences 
should be cancelled completely for physical quantities, as we 
have seen from the coupling constant renormalization. We expect 
that the problem of mixing divergences may not exist when we 
consider real physical processes.
The second problem is the infinite gluon mass correction. In the
time-ordered perturbation theory dimensional regularization is not
available to avoid the nonzero gluon mass correction.  To have
a massless gluon in perturbation theory, we have to introduce
a gluon mass counterterm. In the leading order (one-loop) calculation,
there is no difficulty arising from a gluon mass counterterm.  However, 
when we go to the next order, it has been found that the gluon mass
counterterm leads to a noncancellation of infrared divergences.
The non-vanishing infrared divergences could introduce
non-local counterterms in both the longitudinal and transverse
directions.  In instant quantization, such non-local counterterms
are forbidden for a renormalizable theory.  Here, these
non-local counterterms are allowed by the light-front power
counting. This is a special feature of   LFQCD.  One speculation
from this property is that the non-local counterterms for
infrared divergences may also provide a source for quark confinement
 [\cite{Wilson94}].

In summary, renormalization in {\em   LFQCD Hamiltonian theory}
is very different from conventional Feynman theory and it is an entirely
new subject where investigations are still in their preliminary stage.
In perturbative
calculations, careful treatment could remove all severe infrared
divergences for interesting physical quantities in   LFQCD.
For nonperturbative studies, the cancellation of severe infrared
divergences may not work because certain approximations (e.g.,
Fock space truncation) might be used.  These
approximations may also break many important symmetries such as
gauge invariance and rotational invariance. It is the hope of the
current investigation of light-front renormalization theory
that the counterterms for the light-front
infrared divergences may restore the broken symmetries and
also provide an effective confining   LFQCD Hamiltonian for
hadronic bound states.

\section{Light-Front Heavy Quark Effective Theory (HQET)}

\subsection{About Heavy Quark Symmetry and HQET}
The rich information about electroweak and strong interactions
that can be extracted from various heavy hadron decays has led 
to the extensive exploration of the QCD based and model-independent 
description of heavy hadrons in the past few years. This is 
mainly due to the discovery of heavy quark spin-flavor symmetry 
(HQS) in heavy meson decays by Isgur and Wise  [\cite{Isgur90}]. 
For a typical example, with the HQS, all six form factors in 
$B \rightarrow D$ and $B \rightarrow D^*$ decays are reduced 
to an universal function, called the Isgur-Wise function, 
and the normalization of this universal function at the 
zero-recoil point provides a model-independent determination 
of the Kabayshi-Makawa matrix element $|V_{cb}|$.  Similarly in
heavy-baryon decays, the application of HQS also leads to
tremendous simplifications. 

On the other hand, heavy quark symmetry can be derived from QCD 
in heavy mass limit $m_Q \rightarrow \infty$, via the so-called 
heavy quark effective theory (HQET)  [\cite{Hill}].  The 
later is an effective theory of QCD for heavy quark expansed 
in inverse powers of heavy quark mass $m_Q$.  In fact, HQET 
provides us with a systematical expansion of QCD dynamics in
terms of the dimensionless parameter $\Lambda_{QCD}/m_Q$,
and it serves as a theoretical framework for the systematical 
computation of the $1/m_Q$ corrections to the limit $m_Q 
\rightarrow \infty$. Thus, the HQET offers us a new 
channel to explore the intrinsic properties of hadronic 
structure from QCD.  

In order to actually compute any physical observables and
make definite predictions, one still has to confront the 
non-perturbative QCD dynamics. Currently, except for the 
lattice approach, the main physical quantities, such as  
Isgur-Wise function, can only be computed in various hadronic 
models, such as the constituent quark model, the bag model, 
and QCD sum rules. It would be very interesting if one could 
calculate the Isgur-Wise function, or any hadronic form factors, 
directly from QCD. This requires to construct explicitly the 
heavy hadron bound states within the HQET, which is also 
necessary for  a complete understanding of heavy 
hadron dynamics. We are motivated by such requirement to 
reformulate HQET on the light-front, from which we hope to 
consistently study the heavy hadron bound state problem
 [\cite{zhang95}]. Meanwhile, as we know for light quark
systems, both  quark confinement and spontaneously
chiral symmetry breaking play an essential role to the quark
dynamics in hadrons. In order to provide a nonperturbative 
QCD description for light quark systems, it is necessary to
understand the underlying mechanism for quark confinement
as well as for chiral symmetry breaking. This will certainly
make the problem most complicated. However,
for heavy quark systems, chiral symmetry is explicitly
broken so that confinement is the sole nontrivial
feature influencing heavy quark dynamics. Choosing
the heavy hadron systems should be a good starting point
in the study of nonperturbative QCD. The light-front HQET 
discussed here is mainly based on the works collaborated
with C. Y. Cheung and G. L. Lin  [\cite{zhang95}].

\subsection{$1/m_Q$ Expansion of the Heavy Quark Lagrangian
on the LF}

Let us begin with the QCD Lagrangian for a heavy quark:
\begin{equation} \label{hql}
        {\cal L} = \bar{Q} (i\! \not{\! \! D} - m_Q) Q ,
\end{equation}
where $Q$ is the heavy quark field operator, $m_Q$ the heavy 
quark mass and $D^{\mu}$ the QCD covariant derivative.

In the instant formalism, HQET is obtained by redefining the
heavy quark field as:
\begin{equation}
        Q(x) = e^{-i m_Q v \cdot x} [h_v(x) +H_v(x)]   \label{rdhq1},
\end{equation}
where $v$ is the four velocity of the heavy quark, such that $v^2=1$;
$h_v(x)$ and $H_v(x)$ are respectively the so-called large and small 
components of the heavy quark field, satisfying $\not{\! v}h_v(x)=
h_v(x)$ and $\not{\! v}H_v(x)=-H_v(x)$. From the QCD equation of motion, 
one can express $H_v(x)$ in terms of $h_v(x)$ and show that the former 
is suppressed by $1/m_Q$ compared to the later.  Using Eq.(\ref{rdhq1}) 
and the relation between $h_v(x)$ and $H_v(x)$, one can systematically 
expand the QCD Lagrangian in powers of $1/m_Q$, and arrive at an 
effective theory for the heavy quark.

In the framework of light-front quantization, the situation is
quite different.  Before taking the heavy quark mass limit,
the  quark field is already divided into two parts:
$Q(x) = Q_+(x) + Q_-(x)$,  with $Q_{\pm}(x) = \Lambda_{\pm} Q(x)
={1 \over 2} \gamma^0 \gamma^{\pm} Q(x)$.  The Dirac equation for $Q$
can then be rewritten as two coupled equations for $Q_\pm$:
\begin{eqnarray}
       i D^-Q_+(x) &=& ( i \alpha_{\bot} \cdot D_{\bot}
                + \beta m_Q) Q_- (x),   \label{lffd1}  \\
       i D^+Q_-(x) &=& ( i \alpha_{\bot} \cdot D_{\bot}
                + \beta m_Q) Q_+ (x),   \label{lffd2}
\end{eqnarray}
where $\alpha_{\bot} = \gamma^0 \gamma_{\bot}$ and $\beta = \gamma^0$.  
As we known only the plus-component $Q_+(x)$ is the dynamical field.  
The minus-component $Q_-(x)$ is a light-front constraint that 
can be determined from $Q_+(x)$. In terms of $Q_+(x)$, the QCD 
Lagrangian (1) for the heavy quark can be rewritten as

\begin{equation}
        {\cal L} = Q_+^{\dagger} i D^- Q_+ - Q_+^{\dagger} ( i
                \alpha_{\bot} \cdot D_{\bot} + \beta m_Q ) Q_- ,
\end{equation}
where $Q_-$ can be eliminated by Eq.(\ref{lffd2}).

To derive the light-front HQET, we use the same redefinition 
of the heavy quark field as in the covariant case, 
\begin{equation}
        Q(x) = e^{-i m_Q v \cdot x} {\cal Q}_v(x),  \label{nlfq1}
\end{equation}
but without imposing any constraint on the new variable ${\cal Q}_v$
to separate the large and small components.
It follows that $Q_\pm(x) = e^{-i m_Q v \cdot x} {\cal Q}_{v\pm}(x)$.
Substituting this result into Eq.(\ref{lffd2}), we obtain
\begin{equation}
        {\cal Q}_{v-} (x) = {1 \over m_Q v^+ + iD^+} \Big[i\alpha_{\bot}
                \cdot D_{\bot} + m_Q ( \alpha_{\bot} \cdot v_{\bot} +
                \beta) \Big] {\cal Q}_{v+} (x).
\end{equation}
It is worth noting that in the ordinary light-front formulation of
field theory, the elimination of the dependent component
${Q_-}$ requires the choice of
the light-front gauge $A^+=0$, and a specification of the operator
$1/\partial^+$ which leads to severe light-front infrared problem
that has still not been completely understood [\cite{zhang93a}]. However, 
for the heavy quark field with the redefinition of Eq.(\ref{nlfq1}),
the above problem does not occur since the elimination of the dependent
component ${{\cal Q}_{v-}}$ now depends on the operator $1/(m_Q
v^+ + iD^+)$ which has no infrared problem. Moreover, it has a well 
defined series expansion in powers of $iD^+/m_Q$:
\begin{equation}
        {1 \over m_Q v^+ + iD^+} = {1\over v^+} \, \sum_{n=1}^{\infty} 
		\Big({1 \over m_Q} \Big)^n \Big(-i{D^+\over v^+}\Big)^{n-1}.
\end{equation}
Thus, the heavy quark QCD Lagrangian (\ref{hql}) can be expressed in 
terms of ${\cal Q}_{v+}$ alone. The complete $1/m_Q$ expansion is given 
by 
\begin{eqnarray}
        {\cal L} &=& {1 \over v^+} \Bigg\{2{\cal Q}_{v+}^{\dagger} 
		(iv \cdot D){\cal Q}_{v+} \nonumber \\
	& & ~~~~~~~~~~~ - \sum_{n=1}^{\infty} 
		\Big({ 1 \over m_Q}\Big)^n {\cal Q}_{v+}^{\dagger} 
		\Big\{(i\vec{\alpha}\cdot \vec{D}) \Big(-i {D^+\over
		v^+}\Big)^{n-1} (i \vec{\alpha} \cdot \vec{D}) 
		\Big\} {\cal Q}_{v+} (x) \Bigg\} \nonumber \\
        &=& {\cal L}_0 + \sum_{n=1}^{\infty} {\cal L}_n , \label{lfhqetl}
\end{eqnarray}
where
\begin{equation}
        \vec{\alpha} \cdot \vec{D}= \alpha_{\bot} \cdot D_{\bot} -
                { \alpha_{\bot} \cdot v_{\bot} + \beta \over v^+} D^+ .
\end{equation}
This is the light-front effective heavy quark Lagrangian. 

\subsection{Properties of the Light-Front HQET}

In the symmetry limit, the light-front HQET reduces to
\begin{equation}
        {\cal L}_0 = {2 \over v^+} {\cal Q}_{v+}^{\dagger} (i v \cdot D)
                {\cal Q}_{v+} ,
\end{equation}
which clearly exhibits the flavor and spin symmetries,
because it is independent of Dirac $\gamma$-matrices and
the heavy quark mass, as in the covariant formulation.

However, beyond the symmetry limit, the light-front HQET has 
several advantages over the instant formulation. In the instant 
HQET, the non-leading terms contain high order time-derivatives; 
consequently it is difficult to perform a consistent canonical
quantization beyond the limit $m_Q \rightarrow \infty$  [\cite{Suzuki91}].
It is remarkable to see that in the light-front HQET, only
linear time-derivative appears, and it resides
in ${\cal L}_0$.  The presence of the matrix $ \not{\! n}$ in the
non-leading terms eliminates all light-front time derivative
terms.  This can be seen more clearly in Eq.(\ref{lfhqetl}).  Thus the
canonical quantization of light-front HQET is straightforward:
First of all, the canonical conjugate of the dynamical variable
${\cal Q}_{v+}$ is given by
\begin{equation}
        \Pi_{{\cal Q}_{v+}} = { \partial {\cal L} \over \partial
                (\partial^- {\cal Q}_{v+})} = i {\cal Q}_{v+}^{\dagger},
         \label{cong}
\end{equation}
which does not involve any $1/m_Q$ corrections. Then using the
light-front phase space quantization  [\cite{zhang93a}], we obtain the
basic anti-commutation relation:
\begin{equation}
        \{ {\cal Q}_{v+}(x)~, ~ {\cal Q}^\dagger_{v+}(y) \}_{x^+=y^+} =
                \Lambda_+ \delta^3 (x-y),
\end{equation}
which is valid to all orders in $1/m_Q$.

The second very useful property of the light-front HQET is that the
heavy quark effective Hamiltonian is well defined on the light-front.
From Eqs.(\ref{lfhqetl}) and (\ref{cong}), we obtain the light-front
heavy quark effective Hamiltonian,
\begin{equation}
        H = \int dx^- d^2x_{\bot} {\cal H}(x)
\end{equation}
with the Hamiltonian density ${\cal H}$ given by
\begin{equation}
        {\cal H} = { 1\over iv^+} {\cal Q}^{\dagger}_{v+} (v^-\partial^+
                -2v_{\bot} \cdot \partial_{\bot} ) {\cal Q}_{v+}
                - {2g \over v^+} {\cal Q}^{\dagger}_{v+} (v \cdot A)
                {\cal Q}_{v+} + {\cal H}_{m_Q}
\end{equation}
and
\begin{equation}
        {\cal H}_{m_Q}= \sum_{n=1}^{\infty} {\cal H}_n
                = - \sum_{n=1}^{\infty} {\cal L}_n \, .  \label{lfhqeh}
\end{equation}
This light-front heavy quark effective Hamiltonian can serve as
a basis for constructing heavy hadron bound states, as we will see
in the next lecture. It is also useful for the study of the $1/m_Q$
corrections in heavy quark dynamics. Specifically, suppose we choose 
the light-front gauge ($A^+=0$) in the light-front HQET, we see 
immediately from Eq.(\ref{lfhqetl}) that, in the symmetry breaking 
terms, the power of the gluon field does not increase with that 
of $1/m_Q$. This property, which is unique to the light-front 
formulation, may greatly simplify our treatment of $1/m_Q$ 
corrections. Meanwhile, note that the non-leading
light-front effective Hamiltonian ${\cal H}_n$ is precisely the
minus of the corresponding effective Lagrangian ${\cal L}_n$ given
by Eq.(\ref{lfhqetl}). This simple relation is not valid in
the instant HQET, due to appearance of the high-order
time-derivative terms.

Furthermore, since we have not chosen any specific gauge,
and also there is no light-front infrared divergent problem for
the heavy quark sector, short-distance QCD corrections to the heavy 
quark current and the effective Lagrangian must be the same as those 
calculated in the covariant formulation. Of course, an explicit 
calculation of the short-distance effects in the
light-front HQET is needed to confirm the above statement, which
has not been done as I known.

\subsection{Isgur-Wise Function}

The heavy quark current can also be systematically expanded in
$1/m_Q$ on the light-front.  
In the heavy mass limit, it reduces to the following familiar from:
\begin{equation}
        \overline{Q}^j(x)\Gamma Q^i(x) = e^{-i(m_{Q^j}v'-m_{Q^i}v) 
		\cdot x }\overline{h}_{v}^{jL}(x) \Gamma h_v^{iL} (x) ,
\end{equation}
where $h_v^L = \Big\{1 + {\alpha_\bot \cdot v_\bot + \beta
\over v^+} \Big\} {\cal Q}_{v+}$.
Consequences of the spin symmetry can be readily derived
using this zeroth order heavy quark current.
As an example, consider the matrix elements
\begin{equation}
        \langle P_{Q^j} (v') | \overline{h}_{v'}^{jL} \Gamma
                h_v^{iL} | P_{Q^i} (v) \rangle
                 ~~ {\rm and} ~~ \langle P^*_{Q^j} (v') |
                \overline{h}_{v'}^{jL} \Gamma h_v^{iL} | P_{Q^i} (v)
                \rangle,            \label{27}
\end{equation}
where $\Gamma$ stands for any arbitrary gamma matrix, $P_Q$ and $P^*_Q$
represent respectively a pseudoscalar meson and a vector meson containing
a single heavy quark $Q$. The quantum numbers of the heavy mesons can
be efficiently accounted for by the interpolating fields: $| P_{Q^i} 
(v)\rangle = \overline{h}_v^{iL}\gamma_5 \ell_v | 0 \rangle$,
$| P^*_{Q^i} (v) \rangle = \overline{h}_v^{iL}\not{\!\epsilon}~\ell_v 
| 0 \rangle$, where $\epsilon$ is the polarization vector of the vector 
meson, and $\ell_v$ represents the fully interacting light quark (or 
brown muck). From $\langle 0 | {\cal Q}_{v+} {\cal Q}^{\dagger}_{v+} 
| 0 \rangle = {v^+ \over 2} \Lambda_+$, it is easy to show that
\begin{equation}
        h_v^L \overline{h}_v^L = \Big( 1 + { \alpha_{\bot} \cdot v_{\bot}
                + \beta \over v^+} \Big) {v^+ \over 2} \Lambda_+ \Big( 1
                + { \alpha_{\bot} \cdot v_{\bot} + \beta \over v^+} \Big)
                \beta  = { 1+ \not{\! v} \over 2}.
\end{equation}
Hence, in the heavy mass limit, the heavy meson decay matrix elements
on the light-front take the familiar forms:
\begin{eqnarray}
        & & \langle P_{Q^j} (v') | \overline{h}_{v'}^{jL} \Gamma
                h_v^{iL} | P_{Q^i} (v) \rangle
                = Tr\Big\{ \gamma_5 \Big( {1+ \not{\! v}'
                \over 2} \Big)\Gamma \Big( {1+\not{\! v} \over 2} \Big)
                \gamma_5 M \Big\} \\
        & & \langle P^*_{Q^j} (v') |
                \overline{h}_{v'}^{jL} \Gamma h_v^{iL} | P_{Q^i} (v)
                \rangle  =
                 Tr\Big\{ \not{\! \epsilon}^* \Big( {1 + \not{\! v}'
                \over 2} \Big) \Gamma \Big( {1+ \not{\! v} \over 2}
                \Big) \gamma_5  M \Big\}.
\end{eqnarray}
where M is the transition matrix element for the light quark  [\cite{Wise}],
\begin{equation}
        M = \langle 0 | \overline{\ell}_{v'} \ell_v | 0 \rangle
                \rightarrow  \xi (v' \cdot v) I.
\end{equation}
Thus spin symmetry implies that the transition matrix elements (\ref{27}) 
are described by a single form factor $\xi(v \cdot v')$, which is just 
the famous Isgur-Wise function. An explicit calculation of the Isgur-Wise 
function from light-front bound state wave function will be given in 
the next lecture. 

\section{Quark Confinement and Heavy Hadron Bound States}

\subsection{A Weak-Coupling Treatment to Nonperturbative QCD}

There are two fundamental problems in QCD for hadronic physics, the
quark confinement and the spontaneous breaking of chiral symmetry. 
These two problems are the basis for solving the low-energy hadronic 
bound states from QCD but none of them has been completely understood.
Recently, Wilson et al. proposed a new approach to determine hadronic 
bound states from nonperturbative QCD on the light-front with a 
weak-coupling treatment (WCT)  [\cite{Wilson94}]. The key to eliminating 
necessarily nonperturbative effects is to construct an 
effective QCD Hamiltonian in which quarks 
and gluons have nonzero constituent masses rather than the 
zero masses of the current picture.  The use of constituent
masses cuts off the growth of the running coupling constant
and makes it conceivable that the running coupling never
leaves the perturbative domain. The WCT approach 
potentially reconciles the simplicity of the constituent quark
model with the complexities of QCD. The penalty for achieving 
this weak-coupling picture is the necessity of formulating 
the problem in light-front coordinates and of dealing with 
the complexities of renormalization. 

Succinctly, this new approach of achieving a QCD description of 
hadronic bound states can be summarized as follows: Using a new 
renormalization scheme, called similarity renormalization 
group (SRG) scheme that is recently proposed by Glazek and 
Wilson  [\cite{Glazek94,Wilson94}], one can obtain an effective
QCD Hamiltonian $H_\lambda$ which is a series of expansion in 
terms of the QCD coupling constant, where $\lambda$ is a low 
energy scale. Then one may solve from $H_\lambda$  the 
strongly interacting bound states as a weak-coupling problem.
The WCT scheme contains the following steps: (i) Compute 
explicitly from SRG the $H_\lambda$ up to the second order 
and denote it by $H_{\lambda 0}$ as a nonperturbative part 
of $H_\lambda$.  The remaining higher order contributions in 
$H_\lambda$ are considered as a perturbative part $H_{\lambda I}$. 
(ii) Introduce a constituent picture which allows one to start 
the hadronic bound states with the valence constituent Fock space.  
 The constituent quarks and gluons have masses of 
a few hundreds MeV, and these masses are functions of the scale
$\lambda$ that must vanish when the effective theory goes back 
to the high energy region. (iii) Solve hadronic bound states 
with $H_{\lambda 0}$ nonperturbatively in the constituent picture and 
determine the scale dependence of the constituent masses and the 
coupling constant. The coupling constant $g$ now becomes an effective 
one, denoted by $g_\lambda$. If we could show that with a suitable 
choice of $\lambda$ at the hadronic mass scale, the effective
coupling constant $g_\lambda$ can be arbitrarily small, then 
WCT could be applied to $H_\lambda$ such that the corrections 
from $H_{\lambda I}$ can be truly computed perturbatively. 
If everything listed above works well, we may arrive at a 
weak-coupling QCD theory of the strong interaction for 
hadronic bound states. 

With the idea of SRG and the concept of coupling coherence
 [\cite{Perry93}], Perry has shown that upon a calculation to the 
second order, there exists a logarithmic confining potential in 
the resulting   LFQCD effective Hamiltonian  [\cite{Perry94}]. 
This is a crucial finding to light-front nonperturbative QCD.  
However, the general strategy of solving hadrons through the WCT 
scheme is far to be completed.  Very recently, I used SRG to 
analytically derive from the light-front HQET  [\cite{zhang95}] 
a heavy quark QCD Hamiltonian which is responsible to heavy 
hadron bound states. The resulting 
Hamiltonian explicitly contains a confining interaction between 
a heavy quark and a heavy antiquark at long distance plus a 
Coulomb-type interaction at short distance. With 
this effective QCD Hamiltonian, I study the strongly 
interacting heavy hadronic bound states, from which I can provide 
a WCT to nonperturbative QCD on the light-front, at least for heavy
quarkonia. The following discussion
is mainly based on my recent work, Ref. [\cite{zhang96}].

\subsection{Light-front Similarity Renormalization Scheme}

The basic idea of the SRG approach is to develop a sequence 
of infinitesimal unitary transformations that transform 
an initial bare Hamiltonian $H^B$ to an effective Hamiltonian
$H_\lambda$ in a band-diagonal form relative to an arbitrarily
chosen energy scale $\lambda$:
\begin{equation}
	H_\lambda = S_\lambda H^B S_\lambda^\dagger. \label{st1}
\end{equation}
Here the band-diagonal form means that the matrix elements of 
$H_\lambda$ involving energy jumps 
much larger than $\lambda$ will all be zero, while matrix elements
involving smaller jumps or two nearby energies remain in $H_\lambda$.
The similarity transformation should satisfy the condition that for
$\lambda \rightarrow \infty, H_\lambda \rightarrow H^B$ and $S_\lambda
\rightarrow 1$.

Here, I shall follow the formulation of SRG developed on the 
light-front  [\cite{Wilson94}]. The effective Hamiltonian we seek is 
$H_\lambda$ with $\lambda$ being of order a hadronic mass ($\sim
1$ GeV). We begin with a given bare Hamiltonian which can be
written by $H^B = H_0 + H_I^B$, where $H_0$ is a bare free 
Hamiltonian and $E_i$ is its eigenvalue. Consider an 
infinitesimal transformation, then Eq.(\ref{st1}) is reduced to
\begin{equation} \label{st2}
	{d H_\lambda \over d \lambda} = [H_\lambda, T_\lambda],
\end{equation}
which is subject to the boundary condition $\lim_{\lambda\rightarrow
\infty} H_\lambda = H^B$. 

To force the  Hamiltonian $H_\lambda$
becoming a band-diagonal form in energy space, we need to specify
the action of the generator operator $T_\lambda$. This can be done
by introducing the scale $\lambda$ with $x_{\lambda ij} = {E_j - E_i
\over E_i + E_j + \lambda}$ into a smearing function $f_{\lambda ij} 
= f(x_{\lambda ij})$ such that when $x < 1/3$, $f = 1$; 
when $x > 2/3$, $f =0$; and $f$ may be a smooth function from 
1 to 0 for $ 1/3 \leq x \leq 2/3$. We can write $H_\lambda=
H_0 + H_{I \lambda}$ because $H_0$ is invariant under transformations.
Then Eq.(\ref{st2}) can be reexpressed as
\begin{eqnarray}
	&& {d H_{\lambda ij} \over d\lambda} = f_{\lambda ij}
	  [H_{I \lambda}, T_\lambda]_{ij} + {d \over d\lambda}
	  (\ln f_{\lambda ij}) H_{\lambda ij}, \nonumber \\
	&& T_{\lambda ij} = {1 \over E_j - E_i} \Bigg\{(1-f_{\lambda ij})
	  [H_{I \lambda}, T_\lambda]_{ij} - {d \over d\lambda}
	  (\ln f_{\lambda ij}) H_{\lambda ij} \Bigg\}.
\end{eqnarray}
Here we have used the notation $A_{ij} = \langle i | A | j \rangle$,
and $|i\rangle$ is an eigenstate of $H_0$.
Since $f(x)$ vanishes when $x\geq 2/3$, one can see that 
$H_{\lambda ij}$ does indeed vanish in the far off-diagonal 
region. It also can be seen that $T_{\lambda ij}$ is zero
in the near-diagonal region. 
The solutions for $H_{I \lambda}$ and $T_\lambda$ are 
\begin{equation}
	H_{I\lambda} = H^B_{I \lambda} + {\underbrace{[H_{I \lambda'},
	 T_{\lambda'}]}}_R~, ~~~~~ T_\lambda = H^B_{I \lambda T}
	+ {\underbrace{[H_{I \lambda'}, T_{\lambda'}]}}_T,
\end{equation}
where $H^B_{I \lambda ij} = f_{\lambda ij} H^B_{I ij}$, $H^B_{I\lambda Tij} 
= -{1\over E_j-E_i} \left({d\, \over d\lambda} f_{\lambda ij}\right) 
H^B_{Iij} $, and 
\begin{eqnarray}
	{\underbrace{X_{\lambda^\prime ij}}}_R &=& - f_{\lambda ij}
		\int_\lambda^\infty d\lambda^\prime X_{\lambda^\prime ij}, 
		\label{2} \\
	{\underbrace{X_{\lambda^\prime ij}}}_T &=& - {1 \over E_j - E_i}
		\Big({d\, \over d\lambda} f_{\lambda ij} \Big)
		\int_\lambda^\infty d\lambda^\prime X_{\lambda^\prime ij}
		+ { 1 - f_{\lambda ij} \over E_j -E_i} X_{\lambda ij}.
		\label{3}
\end{eqnarray}
Finally, one obtains an iterated solution for $H_{\lambda}$,
\begin{eqnarray}
	H_{\lambda} &=& \Bigg( H_0 + H^B_{I\lambda} \Bigg) + 
		\Bigg( {\underbrace{[H^B_{I\lambda^\prime}, 
		H^B_{I\lambda^\prime T}]}}_R \Bigg) \nonumber \\
	 && + \Bigg( {\underbrace{[{\underbrace{[H^B_{I\lambda^{
		\prime\prime}}, H^B_{I\lambda^{\prime\prime} 
		T}]}}_{R^\prime} , H^B_{I\lambda^\prime T}]}}_R 
		+{\underbrace{[H^B_{I\lambda^\prime}, 
		{\underbrace{[H^B_{I\lambda^{\prime\prime}}, 
		H^B_{I\lambda^{\prime\prime} T}]}}_{T^\prime}]}}_R \Bigg)
		+ \dots 	\nonumber \\ 
	& = & H^{(0)}_{\lambda}+H^{(2)}_{\lambda}+H^{(3)}_{\lambda}+\dots ,
		\label{eh1}
\end{eqnarray}
Thus, through SRG, we eliminate the interactions
between the states well-separated in energy and generate 
the effective Hamiltonian of eq.(\ref{eh1}). The expansion of 
eq.(\ref{eh1}) in terms of the interaction coupling
constant brings in order by order the full theory corrections 
to this band diagonal low energy Hamiltonian.

Explicitly, the bare Hamiltonian $H^B$ input in the above formulation
can be obtained from the canonical Lagrangian with a high energy 
cutoff that removes the usual UV divergences.  For   LFQCD 
dynamics, the bare Hamiltonian has been constructed in lecture II
(for detailed discussion, see  [\cite{zhang93a,zhang93b}]).
Instead of the cutoff on the field operators which is introduced 
in ref. [\cite{Wilson94}], I shall use a vertex cutoff
to every vertex in the bare Hamiltonian  [\cite{zhang96}]:  
\begin{equation}
	\theta(\Lambda^2/P^+ - |p_i^- - p_f^-|),   \label{ctf}
\end{equation} 
where $p_i^-$ and $p_j^-$ are the initial and final state 
light-front energies respectively between the vertex, 
$\Lambda$ is the UV cutoff parameter, and $P^+$ the 
total light-front longitudinal momentum of the system 
we are interested in.  Eq.(\ref{ctf}) is also called the local 
cutoff in light-front perturbative QCD [\cite{Brodsky81}].
All the $\Lambda$-dependences in the final 
bare Hamiltonian are removed by the counterterms.
The use of eq.(\ref{ctf}) largely simplifies the analysis on 
the cutoff scheme in ref. [\cite{Wilson94}].

Meanwhile, in SRG calculation, we should also give an explicit 
form of the smearing function $f_{\lambda ij}$.  
One of the simplest smearing functions that satisfies 
the requirements of SRG is a theta-function  [\cite{zhang96}]:
\begin{equation}
	f_{\lambda ij} = \theta ({1\over 2} - x_{\lambda ij}). \label{sm1}
\end{equation}
On the light-front, it is convenient to redefine $x_{\lambda ij} =
{|P_i^- - P_j^-| \over P_i^- + P_j^- + \lambda^2/P^+}$. Then we can 
further replace the above smearing function by the following form:
\begin{equation}
	f_{\lambda ij} = \theta({\lambda^2 \over P^+} - |\Delta P_{ij}^-| ), 
		\label{sm2}
\end{equation}
where $\Delta P_{ij}^- = P_i^- - P_j^-$
is the light-front free energy difference between the initial and final 
states of the physical processes.  The light-front free energies of the 
initial and final states are defined as sums over the light-front 
free energies of the constituents in the states. 

With the definition of (\ref{sm2}),  Eq.(\ref{eh1}) can be reduced to
\begin{eqnarray}	
	H_{\lambda ij} &=& \theta({\lambda^2 \over P^+} - |\Delta 
		P_{ij}^-| ) \left\{ H_{ij}^B + \sum_k H^B_{Iik} 
		H^B_{Ikj} \Big[ \frac{g_{\lambda jik}}{\Delta P^-_{ik}}
		+ \frac{g_{\lambda ijk}}{\Delta P^-_{jk}} \Big]
		+ \cdots \right\} .  \label{eh2}
\end{eqnarray}
The front factor 
(the theta-function) in the above equation indicates that  
$H_\lambda$ only describes long distance interactions (with 
respect to the scale $\lambda$) which is responsible to
hadronic bound states.  The function $g_{\lambda ij}$ 
in eq.(\ref{eh2}) is given by
\begin{eqnarray}
	g_{\lambda ijk} &=& \int_{\lambda^2/P^+}^\infty d({\lambda'}^2/ P^+)
		f_{\lambda' ik} {d \over d({\lambda'}^2/P^+)}
		f_{\lambda' jk} \nonumber \\
	 &=& \theta(|\Delta P_{jk}^-|-{\lambda^2/P^+})
		\theta( |\Delta P_{jk}^-| - |\Delta P_{ik}^-|) .
		\label{sd}
\end{eqnarray}

\subsection{Heavy Quark Confining Interaction}
 
Now we can use SRG to the light-front HQET to derive a heavy quark 
confining Hamiltonian, from which we may solve from QCD the heavy 
hadron bound states directly.

In the large $m_Q$ limit, only the leading (spin and mass 
independent) Hamiltonian is remained. The $1/m_Q^n$ terms 
($n \geq 1$) in (\ref{lfhqetl}) can 
be regarded as perturbative corrections to the leading order 
operators and states. To determine 
confining interactions in heavy quark systems, the leading 
heavy quark Hamiltonian plays an essential role.  
With the light-front gauge $A^+=0$, the leading-order 
bare QCD Hamiltonian density is 
\begin{eqnarray}
    {\cal H}_{ld} &=& { 1\over iv^+} {\cal Q}^{\dagger}_{v+} (v^-\partial^+
                -2v_{\bot} \cdot \partial_{\bot} ) {\cal Q}_{v+} \nonumber \\
	& & ~~~~~~ 
     		- {2g \over v^+} {\cal Q}^{\dagger}_{v+} \left\{v^+ \Big[
		\Big({1\over \partial^+}\Big) \partial_{\bot} \cdot A_{\bot}
		\Big] - v_{\bot} \cdot A_{\bot} \right\} {\cal Q}_{v^+}
		\nonumber \\
     & & ~~~~~~ + 2g^2 \Big({1\over \partial^+}\Big) \Big({\cal 
		Q}^{\dagger}_{v+} T^a {\cal Q}_{v^+} \Big) \Big({1 \over
		\partial^+} \Big) \Big( \psi_+^\dagger T^a 
		\psi_+ \Big) , \label{ldhh}
\end{eqnarray}
where $\psi_+$ is either the heavy antiquark field or the 
light-front quark field operator in the present consideration. 
Note that besides the leading term in eq.(\ref{lfhqetl}), 
the above bare Hamiltonian has also already included 
the relevant terms from the gauge field part, $-{1\over 2}
{\rm Tr} (F_{\mu\nu} F^{\mu\nu})$, of the QCD Lagrangian.  
These terms come from the elimination of the unphysical gauge 
degrees of freedom, the longitudinal component $A^-_a$ [\cite{zhang93b}].
Eq.(\ref{ldhh}) has obviously the spin and flavour heavy
quark symmetry, or simply the heavy quark symmetry.

The above leading Hamiltonian (or Lagrangian) is the basis 
of the QCD-based description for heavy hadrons containing a single 
heavy quark, such as $B$ and $D$ mesons.  As recently pointed 
out by Mannel et al. [\cite{Mannel95}] the purely heavy
quark leading Lagrangian may be not appropriate to describe
heavy quarkonia.  This is because the 
anomalous dimension of QCD radiative correction to 
$Q\overline{Q}$ currents contains an infrared singularity 
in the limit of two heavy constituents having equal
velocity. Such an infrared singularity is a long distance 
effect and should be absorbed into quarkonium states.  
To avoid this problem, they argued that one may incorporate 
the effective Hamiltonian with at least the first order kinetic 
energy term into the leading Hamiltonian  [\cite{Mannel95}]. The
light-front kinetic energy can be obtained from eq.(\ref{lfhqeh}),
\begin{equation}
	{\cal H}_{kin} = - {1\over m_Q v^+} {\cal Q}_{v+}^\dagger
		\Bigg\{ \partial_\bot^2 - {2 v_\bot \cdot 
		\partial_\bot \over v^+}
		\partial^+ + {v^- \over v^+} \partial^{+2} \Bigg\}
		{\cal Q}_{v+}.	\label{nloh}
\end{equation}
As a consequence, in the heavy mass limit, quarkonia have 
spin symmetry but no flavour symmetry. 

i). {\it Confining Hamiltonian for Heavy Quarkonia}.
Within light-front HQET, we now follow the procedure 
described above to find an effective QCD 
Hamiltonian for $Q\overline{Q}$ systems. The bare Hamiltonian 
for $Q\overline{Q}$ systems contains (\ref{ldhh}) and (\ref{nloh}) 
for both heavy quark and antiquark plus the full QCD Hamiltonian
for gluons and light quarks [\cite{zhang93b}]. Since the kinetic 
energy (\ref{nloh}) should be the same order of the Coulomb 
interaction, we may treat the kinetic energy in the same way 
as the instantaneous $Q\overline{Q}$ interaction 
[the last term in eq.(\ref{ldhh})]. Thus, the free Hamiltonian 
$H_0$ used in SRG is given only by the first term in eq.(\ref{ldhh}) 
plus the free gluon Hamiltonian.

With the above consideration, it is easy to compute the
effective Hamiltonian eq.(\ref{eh2}) for $Q\overline{Q}$ systems.
Following the WCT ideas, we shall  
calculate $H_\lambda$ for $Q\overline{Q}$ systems up to the 
second order in the initial and final states 
defined by $| i \rangle = b^{\dagger}_v(k_1, \lambda_1) 
d^{\dagger}_v(k_2,\lambda_2) | 0 \rangle$ and $| j \rangle = 
b^{\dagger}_v(k_3, \lambda_3) d^{\dagger}_v(k_4,\lambda_4) 
| 0 \rangle$, respectively, where $k_i$ is the residual momentum 
of heavy quarks, $p^{\lambda}_i= m_Q v^{\lambda} + k^{\lambda}_i$, 
and $\lambda_i$ its helicity.  The result is 
\begin{equation}
	H_{\lambda 0 ij} = H_{Q\overline{Q}free ij}  
		+ V_{Q\overline{Q} ij}  , \label{QQeh}
\end{equation}
where
\begin{eqnarray}
	H_{Q\overline{Q}free ij} &&= [2(2\pi)^3]^2 \delta^3
		(\bar{k}_1-\bar{k}_3) \delta^3(\bar{k}_2-\bar{k}_4) 
	 	\delta_{\lambda_1 \lambda_3} \delta_{\lambda_2 
		\lambda_4} \nonumber \\
	& & \times \Bigg\{ {\overline{\Lambda}\over m_Q}
	 	\Big[2\kappa_\bot^2 + \overline{\Lambda}^2(2y^2-2y +1)
		\Big] - \overline{\Lambda}^2 - 2{g^2 \over 4\pi^2} 
		C_f {\lambda^2 \over K^+}\ln \epsilon \Bigg\} , 
			\label{QQfe} \\
	V_{Q\overline{Q} ij}&& (y-y', \kappa_\bot -\kappa'_\bot) 
	     =  2(2\pi)^3 \delta^3(\bar{k}_1 + \bar{k}_2 - \bar{k}_3 - 
		\bar{k}_4) \delta_{\lambda_1 \lambda_3} \delta_{\lambda_2 
		\lambda_4}  \nonumber \\
	&& \times {-4g^2 (T^a)(T^a) \over (K^+)^2} \Bigg\{{1 
		\over (y-y')^2} \Big(1 - \theta( A(y-y', \kappa_\bot 
		- \kappa'_\bot, \overline{\Lambda}) - \lambda^2) \Big)
		\nonumber \\ &&
	  + {\overline{\Lambda}^2 \over (\kappa_{\bot} - 
		\kappa_{\bot}')^2 + (y-y')^2 \overline{\Lambda}^2} 
		\theta(A(y-y', \kappa_\bot - \kappa'_\bot, 
		\overline{\Lambda}) - \lambda^2) 
		\Bigg\}  \label{QQvv}.
\end{eqnarray}
Here we have introduced the longitudinal residual momentum 
fractions and the relative transverse residual momenta, 
\begin{eqnarray}
	&& y = k^+_1 / K^+ ~,~~~~~ \kappa_{\bot} = 
		k_{1\bot} - y K_{\bot} , \nonumber \\
	&& y' = k^+_3 / K^+ ~,~~~~~ \kappa'_{\bot} = 
		k_{3\bot} - y' K_{\bot},	\label{QQlmf}
\end{eqnarray}
where $K^\mu$ is defined as the residual center mass momentum 
of the heavy quarkonia: $K^\mu = \overline{\Lambda} v^{\mu}$,
and $\overline{\Lambda} = M_H - m_{Q} - m_{\overline{Q}}$ is a 
residual heavy hadron mass. It follows that $K^+ = k_1^+ + k_2^+ 
= k_3^+ + k_4^+$, $K_{\bot}= k_{1\bot} + k_{2\bot} = k_{3\bot} 
+ k_{4\bot}$. Since  $ 0 \leq p_1^+ = m_Q v^+ + k_1^+ \leq M_Hv^+$,
in the heavy quark mass limit, we have $M_H \rightarrow 2m_Q$ 
so that $-m_Qv^+ \leq k^+_1, ~ k^+_3 \leq m_Q v^+$.  Hence, 
the range of $y$ and $y'$ are given by $- \infty < y , y' < \infty $.
We have also defined in eq.(\ref{QQvv})
\begin{equation}
	A(y-y', \kappa_\bot - \kappa'_\bot, \overline{\Lambda}) 
		\equiv {(\kappa_\bot - \kappa'_\bot)^2 \over |y-y'|} 
		+ |y-y'| \overline{\Lambda}^2 .
\end{equation}

Eq.(\ref{QQeh}) is the nonperturbative part of the effective 
Hamiltonian for heavy quarkonia in the WCT scheme, in which 
we have already let UV cutoff parameter $\Lambda \rightarrow 
\infty$ and the associated divergence has been put in the mass 
correction. The kinetic energy (\ref{nloh}) is now included 
in the above effective Hamiltonian [the $1/m_Q$ term in 
eq.(\ref{QQfe})]. Note that there is an infrared divergent 
term in eq.(\ref{QQfe}) which comes from the quark self-energy 
correction in SRG,
where $\epsilon$ is an infrared cutoff of the momentum fraction
$q^+/K^+$, and $q^+$ the longitudinal momentum carried by gluon
in the quark self-energy loop.  The usual mass correction 
$\delta m_Q^2 = {g^2 \over 4\pi^2} C_f \overline{\Lambda}^2\ln 
{\Lambda^2 \over \lambda^2}$, has been 
renormalized away in eq.(\ref{QQfe}). In the WCT scheme, by removing 
away this mass correction, we should assign the corresponding
constituent quark mass in $H_{\lambda 0}$ being $\lambda$-dependent.  
But, the heavy quark mass is larger than the low energy scale.  
Its dependence on $\lambda$ should be very weak and could be 
neglected. While, the $Q\overline{Q}$ interaction (\ref{QQvv})
contains two contributions: the instantaneous interaction plus 
the second order contribution in eq.(\ref{eh2}) [i.e.  
the terms proportional to the theta function in eq.(\ref{QQvv})].
We shall show next that the above $V_{Q\overline{Q}}$ is indeed a 
combination of a confining interaction plus a Coulomb-type 
interaction.

ii). {\it Quark Confinement on the Light-Front}.
In our framework,   LFQCD vacuum is trivial. The nature 
of nontrivial QCD vacuum structure, the confinement as well as 
the chiral symmetry breaking, must made manifestly in 
$H_\lambda$ in terms of new
effective interactions. We will see that $H_{\lambda 0}$
explicitly contains a confining interaction at long distances.
The interactions associated with the chiral 
symmetry breaking may be manifested in the fourth order 
computation of $H_\lambda$ for light quark systems
 [\cite{Wilson94}], but these interactions are not important 
in the study of heavy hadrons here. 

The confining interaction can be easily obtained by applying the
Fourier transformation to the first term in (\ref{QQvv}).  
It is convenient to perform the calculation in the frame
$K_\bot =0$, in which
\begin{eqnarray}
	 \int {dq^+ d^2q_\bot \over (2\pi)^2} && e^{i(q^+ x^- 
		+ q_\bot \cdot x_\bot)} \Bigg\{-4g_\lambda^2 
		~{1 \over q^{+2}} \theta(\lambda^2 - A(q^+/K^+, q_\bot, 
		\overline{\Lambda})) \Bigg\}  \nonumber \\
	&&= - {g_\lambda^2 \over 2\pi^2} \int_0^{{\lambda^2 \over 	
		\overline{\Lambda}^2} K^+} dq^+ e^{iq^+ x^-}{q^2_{\bot m} 
		\over q^{+2}}~{2J_1(|x_\bot|q_{\bot m}) \over 
		|x_\bot| q_{\bot m}},		\label{conp}
\end{eqnarray}
where we have used the relation $q^+ = k_1^+-k_3^+=K^+(y-y'), 
q_\bot = k_{1\bot} - k_{3\bot} = \kappa_\bot - \kappa'_\bot$ 
for $K_\bot =0$, while $q_{\bot m} \equiv 
\sqrt{{\lambda^2 \over K^+} q^+ - {\overline{\Lambda}^2 \over 
K^{+2}} q^{+2}}$, and $J_1 (x)$ is a Bessel function.   
An analytic solution to the
integral (\ref{conp}) may be difficult to carry out. 
However, the nature of confining interactions is a large 
distance QCD behavior.  We may consider the integral
for large $x^-$ and $x_\bot$.  In this case, if $q^+ x^-$ and/or 
$|x_\bot|q_{\bot m}$ are large, the integration vanishes,
yet $J_1(x) = {x\over 2} + {x^3\over 16} + \cdots$ for
small $x$.  The dominant contribution of the integral 
(\ref{conp}) for large $x^-$ and $x_\bot$ comes from the small 
$q^+$ such that $q^+ x^-$ and/or $|x_\bot|q_{\bot m}$ must 
remain small, which leads to $e^{iq^+ x^-}{2J_1(|x_\bot|q_{\bot m}) 
\over |x_\bot| q_{\bot m}} \simeq 1 $. This corresponds to 
$q^+ < {1 \over x^-}$ and/or $q^+ < {K^+ \over |x_\bot|^2 
\lambda^2 }$.

If $q^+ < {1 \over x^-}< { K^+ \over |x_\bot|^2 \lambda^2 }$, 
eq.(\ref{conp}) is reduced to
\begin{equation}
	-{g_\lambda^2 \over 2\pi^2} \int_0^{1\over x^-} dq^+  
		{1\over q^{+2}}\Bigg({\lambda^2 \over K^+} q^+ - 
		{\overline{\Lambda}^2 \over K^{+2}} q^{+2} \Bigg) 
	    = {g_\lambda^2\lambda^2\over 2\pi^2 K^+} 
		 \Big(\ln (K^+|x^-|) + \ln \epsilon \Big) , \label{conp1}
\end{equation}
where a term $\sim {1 \over x^-}$ is neglected since $x^-$ is large,
and $\epsilon$ is an infrared cutoff of the momentum fraction 
$q^+/K^+$. It is the same as the divergence occurs in the quark
self-energy contribution so that the above
 infrared logarithmic divergence ($\sim \ln \epsilon$) exactly 
cancels the divergence in eq.(\ref{QQfe}) for color single states.
What remains is a logarithmic confining interaction 
(except for a color factor):
\begin{equation}
	V_{conf.}(x^-, x_\bot) \sim {g_\lambda^2 \lambda^2\over 
		2\pi^2 K^+} \ln (K^+|x^-|).
\end{equation}
Similarly, when  $q^+ < { K^+ \over |x_\bot|^2 \lambda^2 }< 
{1 \over x^-}$, we have
\begin{equation}
	-{g_\lambda^2 \over 2\pi^2} \int_0^{K^+ \over 
		|x_\bot|^2 \lambda^2} dq^+ {1\over q^{+2}}\Big({\lambda^2
		\over K^+} q^+ - {\overline{\Lambda}^2 \over 
		K^{+2}} q^{+2} \Big) = {g_\lambda^2 \lambda^2\over 
		2\pi^2 K^+} \Big(\ln ( \lambda^2 |x_\bot|^2 ) + 
		\ln \epsilon \Big) ,	\label{conp2}
\end{equation}
where the term $\sim {1 \over x^2_\bot}$ has also been ignored 
because of the large $x^2_\bot$. Again, the infrared divergence 
($\sim \ln \epsilon$) is cancelled in $H_\lambda$ for physical
states, and we obtain the following confining interaction:
\begin{equation}
	V_{conf.}(x^-, x_\bot) \sim {g_\lambda^2 \lambda^2\over 
		2\pi^2 K^+} \ln (\lambda^2 |x_\bot|^2).
\end{equation}
Hence, the effective Hamiltonian $H_{\lambda 0}$ exhibits a logarithmic 
confining interaction between a heavy quark and a heavy antiquark 
in all the directions of $x^-$ and $x_\bot$ space.  

The Coulomb interaction corresponds the second term in (\ref{QQvv}),
its Fourier transformation (except for the color factor) is
\begin{equation}
	{\overline{\Lambda}^2  \over (\kappa_{\bot} - 
		\kappa_{\bot}')^2 + (y-y')^2 \overline{\Lambda}^2} 
		\sim  {1 \over 4\pi} \int dx^- d^2x_\bot e^{i(x^-q^+
		+ q_\bot \cdot x_\bot)}
		\Bigg({\overline{\Lambda} \over K^+} \Bigg) {1 \over 
		r_l} ,	\label{Coulp}
\end{equation}
where $r_l \equiv 
\sqrt{x^2_\bot + \Big({\overline{\Lambda} \over K^+} \Big)^2 
(x^-)^2}$ which is defined as a ``radial'' variable in the light-front 
space [\cite{Wilson94}]. Eq.(\ref{Coulp}) shows that the Coulomb 
interaction on the light-front has the form
\begin{equation}
	V_{Coul.}(x^-, x_\bot) \sim  - {g_\lambda^2 \over 4\pi} 
		{\overline{\Lambda} \over K^+} {1 \over r_l}.
\end{equation}
Thus, we have explicitly shown that $H_{\lambda 0}$ contains a 
Coulomb interaction at short distances and a confining interaction 
at long distances.
 
Moreover, a clear light-front picture of quark confinement 
emerges here. To be specific, we define quark confinement 
as follows: i) There is a 
confining interaction between quarks such that quarks cannot be 
well-separated; ii) No color non-singlet bound states exist 
in nature, only color singlet states with finite masses can 
be produced and observed; and iii) The conclusions of i--ii)
are only true for QCD but not for QED.  

We have shown explicitly the existence of a confining 
interaction in $H_{\lambda 0}$. One can also easily see 
from $H_{\lambda 0}$ the non-existence of color non-singlet 
bound states.  This is essentially related to the infrared 
divergences in $H_{\lambda 0}$. From eqs.(\ref{conp1}) and 
(\ref{conp2}), we find that the uncancelled instantaneous 
interaction contains a logarithmic infrared divergence.
Except for the color factor, this infrared divergence has the 
same form as the divergence in eq.(\ref{QQfe}).
Thus, we immediately obtain the following conclusions.

	(a). For a single (constituent) quark state, the 
interaction part of $H_{\lambda 0}$ does not contribute to
its energy. The remaining infrared divergence from quark
self-energy correction implies that the dynamical quark
mass for a single quark state is infinite (infrared divergent) 
and cannot be renormalized 
away in the spirit of gauge invariance.  Equivalently speaking,
single quark states carry an infinitely large mass and therefore
they cannot be produced. 

	(b). For color non-singlet composite states, the color 
factor $(T^a)_{\alpha \beta} (T^a)_{\delta \gamma}$ in the 
$Q\overline{Q}$ interaction is different from the color factor 
$C_f=\tr(T^a T^a)$.
Therefore, the infrared divergence in the self-energy correction
also cannot be cancelled by the corresponding divergence
from the uncancelled instantaneous interaction. As a result,
color non-singlet composite states are infinitely heavy
that they cannot be produced as well.

	(c). For color singlet $Q\overline{Q}$ states, the 
color factor $(T^a)(T^a) \rightarrow C_f$.  Thus, 
the infrared divergences are completely cancelled and the resulting
effective Hamiltonian is finite.  In other words, only color 
singlet composite are physically observable.

Finally, we argue that the above mechanism of quark 
confinement is indeed only true for QCD.  As we have
seen the light-front confinement interaction is just an 
effect of the non-cancellation between instantaneous 
interaction and one transverse gluon interaction generated
in SRG. Such a non-cancellation arises in SRG because we 
introduce the energy scale $\lambda$. Introducing the energy
scale $\lambda$ in SRG forces the transverse gluon energy 
involved in the $Q\overline{Q}$ effective interaction 
never be less than a certain value (the energy scale $\lambda$). 
This implies that the gluon may become
massive at the hadronic mass scale.  Of course, such a gluon mass 
must be a dynamical mass generated from the highly nonlinear gluon 
interactions. In other words, the above confining picture is 
indeed a dynamical consequence of non-Abelian gauge theory. This 
confinement mechanism is not valid in QED.  In QED, since  
photon mass is always zero, the photon energy covers the entire 
range from zero to
infinity.  Thus, in QED, we can always choose the energy 
scale $\lambda$ being zero. With $\lambda=0$, the infrared 
divergences do not occur in the electron self-energy 
correction.  As a result, the renormalized single electron mass
is finite, in contrast to the divergent mass of single quark 
states.  For the same reason, with $\lambda=0$, 
the instantaneous interaction in
the effective QED Hamiltonian is also exactly cancelled
by the same interaction from one transverse photon exchange  
so that only one photon exchange Coulomb interaction remains.  
Thus, applying SRG to QED  and let $\lambda =0$ in 
the end of procedure, we obtain a conventional QED 
Hamiltonian which only contains the Coulomb interaction.  

iii). {\it Extension to Heavy-Light Quark Systems}. 
We can also apply SRG to the heavy-light quark system (heavy hadrons 
containing one heavy quark).
The bare cutoff Hamiltonian we begin with for heavy-light 
quark systems is the combination of the heavy quark effective 
Hamiltonian (\ref{ldhh}) and the full Hamiltonian for the light 
quarks and gluons.  We may also introduce
the residual center mass momentum for heavy-light systems,
$K^+ = \overline{\Lambda} v^+ = p^+_1 + k^+_1 = p^+_2 + k^+_2$,
$K_\bot = \overline{\Lambda} v_\bot = p_{1\bot} + k_{1\bot}
 = p_{2\bot} + k_{2\bot}$, 
where $\overline{\Lambda} = M_H-m_Q$, $p_1$ and $p_2$ are 
the light antiquark momenta and $k_1$ and $k_2$ the residual
momenta of the heavy quarks in the initial and final $Q\overline{q}$
states respectively.  

Following the general procedure, it is easy to find the 
nonperturbative part of the effective Hamiltonian for 
heavy-light quark systems,
\begin{equation}
	H_{\lambda 0 ij} = \theta({\lambda^2 \over K^+} -|\Delta 
		K_{ij}^-| ) \Big\{ H_{Q\overline{q}free ij}  
		+ V_{Q\overline{q} ij} \Big\} , \label{Qqeh}
\end{equation}
where
\begin{eqnarray}
	H_{Q\overline{q}free ij} &&= [2(2\pi)^3]^2 \delta^3
		(\bar{k}_1-\bar{k}_2) \delta^3(\bar{p}_1-\bar{p}_2) 
	 	\delta_{\lambda_1 \lambda_3} \delta_{\lambda_2 
		\lambda_4}	 \nonumber \\
	& & ~~~\times \Bigg\{ (y-1) \overline{\Lambda}^2 +
		{\kappa^2_{\bot} + m_q^2 \over y } 
		- {g^2 \over 2\pi^2 } C_f {\lambda^2 \over K^+}
		\ln \epsilon \Bigg\} , \label{Qqeh1} \\
	V_{Q\overline{q} ij}&& (y-y',\kappa_\bot - \kappa'_\bot)=2(2\pi)^3 
		\delta^3(\bar{k}_1 + \bar{p}_1 - \bar{k}_2 - \bar{p}_2) 
		\delta_{\lambda_1 \lambda_3} \delta_{\lambda_2 \lambda_4} 
		\nonumber \\ 
	&& ~\times {- 2g^2 (T^a)(T^a) \over (K^+)^2} \Bigg\{ {2\over (y-y')^2}
		- \Bigg[ 2{(\kappa_\bot-\kappa'_\bot)^2 \over 
		(y - y')^2} - {\kappa^2_\bot -\kappa_\bot \cdot 
		\kappa'_\bot \over y(y-y') } \nonumber \\
	&& ~~ - {\kappa_\bot \cdot
		\kappa'_\bot - (\kappa')^2_\bot \over y'(y-y')} 
		\Bigg] \Bigg[ {\theta(B - \lambda^2) 
		\theta(B - A) \over (\kappa_\bot-\kappa'_\bot)^2 
		-(y-y')( {\kappa_\bot^2 \over y} - {(\kappa')_\bot^2
		\over y'})} \nonumber \\
	& & ~~~~~~~ + {\theta(A - \lambda^2) \theta(A - B) \over
		(\kappa_\bot-\kappa'_\bot)^2 + (y-y')^2 \overline{
		\Lambda}^2} \Bigg] \Bigg\}, \label{Qqvv}
\end{eqnarray} 
with  $B \equiv \Bigg|{(\kappa_\bot-\kappa'_\bot)^2 \over y-y'}
 - {\kappa_\bot^2 \over y} + {(\kappa')_\bot^2 \over y'} \Bigg| $
and the function $A$ has the same form as in quarkonium case.  Here 
we have also introduced $ y = p^+_1 / K^+ $, $\kappa_{\bot} = p_{1\bot} 
- y K_{\bot}$, but the range of $y$ is now given by $ 0 < y 
= {M_H \over \overline{\Lambda}} {p_1^+ \over P^+} < \infty $.

The heavy-light quark effective Hamiltonian is $m_Q$-independent. 
This is because in heavy-light quark systems the heavy quark 
kinetic energy can be treated as a perturbative correction to 
$H_{\lambda 0}$. Obviously the above $H_{\lambda 0}$ 
has the heavy quark spin and flavour symmetry.  Compared 
to the $V_{Q\overline{Q}}$, $V_{Q\overline{q}}$ interactions 
are much more complicated. But it is not difficult to check that 
the above $V_{Q\overline{q}}$ contains a confining interaction.  The
confining mechanism is the same for $Q\overline{Q}$ and 
$Q\overline{q}$ systems, as well as for $q\overline{q}$
systems, as one can show.

In conclusion, we have obtained the nonperturbative part of a 
confining QCD Hamiltonian for heavy-heavy and heavy-light quark 
systems.  We are now ready to solve heavy 
hadron states on the light-front and to show how the WCT
scheme works in the present formulation

\subsection{Heavy Hadron Bound States}

As we mentioned the ideas of WCT to nonperturbative QCD
is to begin with the effective QCD Hamiltonian $H_\lambda = 
H_{\lambda 0} + H_{\lambda I}$. Then using the constituent picture
to solve nonperturbatively the hadronic bound state equations
governed by $H_{\lambda 0}$ and to determine the running coupling 
constant $g_\lambda$.  If one could properly choose the nonperturbative 
$H_{\lambda 0}$ such that $g_\lambda$ is arbitrarily 
small, then the corrections from $H_{\lambda I}$ could be computed 
perturbatively, and we would say that a WCT to nonperturbative 
QCD is realized. Now, I shall discuss such a WCT to heavy hadron 
bound states.

i). {\it Heavy Hadron Bound State Equation Under WCT.}
As we have pointed out in the first lecture solving eq.(\ref{bse}) 
from QCD with the entire Fock space is impossible.  A basic 
motivation of introducing the WCT scheme is to simplify the 
complexities in solving the above equation. In the present 
framework, $H_{LF}= H_\lambda$,
where $H_\lambda$ has already decoupled from high energy states.  
Furthermore, the reseparation $H_\lambda=H_{\lambda 0} + 
H_{\lambda I}$ is another crucial step in WCT, where only
$H_{\lambda 0}$ is assumed to have the nonperturbative 
contribution to bound states through eq.(\ref{bse}), 
and $H_{\lambda I}$ is supposed to be a perturbative term 
which should not be considered when we try to solve
eq.(\ref{bse}) nonperturbatively.  

The next important step in the WCT scheme is the use of a 
constituent picture. The success of the constituent quark 
model suggests that we may only consider the valence 
quark Fock space in determining the hadronic bound 
states from $H_{\lambda 0}$.  In this picture, 
quarks and gluons must have constituent masses. 
This constituent picture can naturally be realized on
the light-front  [\cite{Wilson94}]. However, an essential 
difference from the phenomenological constituent 
quark model description is that the constituent masses 
introduced here are $\lambda$ dependent.  The scale
dependence of constituent masses (as well as the effective
coupling constant) is determined by solving the bound 
states equation and fitting the physical quantities with 
experimental data. But for heavy quark mess, this 
$\lambda$-dependence can be ignored.
Once the constituent picture is introduced,
we can truncate the general expression of the light-front 
bound states to only including the valence quark Fock 
space.  The higher Fock space contributions can be 
recovered as a perturbative correction through $H_{\lambda I}$.  
Thus, eq.(\ref{wf1}) for heavy quarkonia can be approximately
written as:
\begin{eqnarray}
        | \Psi (K^+, K_\bot, \lambda_s) \rangle = \sum_{\lambda_1 
		\lambda_2} \int && [d^3\bar{k}_1] [d^3\bar{k}_2]  
		2(2\pi)^3 \delta^3(\bar{K} - \bar{k}_1-\bar{k}_2) 
			\nonumber \\
		&& \times \phi_{Q\overline{Q}}
         	(y,\kappa_{\bot}) b^\dagger_v(k_1,\lambda_1) 
		d^\dagger_{-v}(k_2, \lambda_2) |0 \rangle ,   \label{QQwf}
\end{eqnarray}
where the wavefunction $\phi_{Q\overline{Q}}(y,\kappa_{\bot})$
may be mass dependent due to the kinetic energy in $H_{\lambda 0}$
[see (\ref{QQfe})] but it is spin independent in heavy mass limit. 
Also note that the heavy quarkonium states in heavy mass limit
are labelled by the residual center mass momentum $K^\mu$.
We may normalize eq.(\ref{QQwf}) as follows:
\begin{equation}
        \langle \Psi(K'^+,K'_\bot,\lambda_s') | \Psi(K^+,K_\bot,
		\lambda_s) \rangle = 2(2\pi)^3 K^+ \delta^3
		(\bar{K}-\bar{K}') \delta_{\lambda'_s \lambda_s}, 
		\label{nQQbs}
\end{equation}
which leads to 
\begin{equation}
       \int {dy d^2\kappa_{\bot} \over 2 (2\pi)^3} |\phi_{Q\overline{Q}}
		(y,\kappa_{\bot})|^2 = 1. \label{nQQwf}
\end{equation} 

With the above analysis on the quarkonium states, 
it is easy to derive the corresponding bound state equation.  
Let  $H_{LF} = H_{\lambda 0}$ of eq.(\ref{QQeh}), 
eq.(\ref{lfbse}) is reduced to 
\begin{eqnarray}
     \Bigg\{2\overline{\Lambda}^2 - {\overline{\Lambda}\over m_Q}
	&& \Big[2\kappa_\bot^2 + \overline{\Lambda}^2(2y^2-2y +1)
	\Big] \Bigg\} \phi_{Q\overline{Q}}(y,k_{\bot}) \nonumber \\
	=&& \Bigg(-{g_\lambda^2\over 2 \pi^2} \lambda^2 C_f \ln\epsilon\Bigg)
		~\phi_{Q\overline{Q}}(y,k_{\bot}) \nonumber \\
	&& ~ -4g_\lambda^2 (T^a)(T^a) \int {dy' d^2\kappa'_{\bot} \over 
		2(2\pi)^3} \Bigg\{ {1 \over (y-y')^2} \theta( 
		\lambda^2 - A) \nonumber \\ && ~~~
	  +  {\overline{\Lambda}^2 \over (\kappa_{\bot} - 
	  \kappa_{\bot}')^2 + (y-y')^2 \overline{\Lambda}^2} \theta
		(A - \lambda^2) \Bigg\} \phi_{Q\overline{Q}}
		(y',\kappa'_{\bot}).  \label{QQbse}
\end{eqnarray}
This is the light-front bound state equation for heavy quarkonia 
in the WCT scheme. 

For the heavy mesons containing one heavy quark, similar consideration
leads to
\begin{eqnarray}
     \Big(\overline{\Lambda}^2 + && (1-y) \overline{\Lambda}^2 
	- {\kappa_\bot^2 + m^2_q(\lambda) \over y} \Big)
	\Phi_{Q\overline{q}}(y,k_{\bot},\lambda_1, 
	\lambda_2) \nonumber \\
    = && \Big(-{g_\lambda^2\over 2 \pi^2} \lambda^2 C_f \ln\epsilon\Big)
		~\Phi_{Q\overline{q}}(y,k_{\bot},
		\lambda_1, \lambda_2) \nonumber \\
	&& ~~ + (K^+)^2\int {dy' d^2\kappa'_{\bot} \over 
		2(2\pi)^3} V_{Q\overline{q}}(y-y',\kappa_{\bot}
		- \kappa_{\bot}')\Phi_{Q\overline{q}}
		(y',\kappa'_{\bot}, \lambda_1, \lambda_2),  
		\label{Qqbse}
\end{eqnarray}
where $V_{Q\overline{q}}$ is given by eq.(\ref{Qqvv}). Note 
that the light antiquark here is a brown muck, a current light 
antiquark surround by infinite gluons and $q\overline{q}$ 
pairs that results in a constituent quark mass $m_q$ which is
a function of $\lambda$.

ii). {\it A General Analysis of Light-Front Wavefunctions.}
A numerical computation to the bound state equations,
eqs.(\ref{QQbse}) and (\ref{Qqbse}), is actually not too difficult.
However, to have a deeper insight about the internal structure of
light-front bound states, it is better to have
an analytic analysis.  For this propose, we would like to present
a general analysis of light-front hadronic wavefunctions.

The heavy hadronic wavefunctions in the heavy mass limit 
are rather simple. 
First of all, the heavy quark kinematics have already added 
some constraints on the general form of the light-front
wavefunction $\phi (x, \kappa_\bot)$.  When we introduce the 
residual longitudinal
momentum fraction $y$ for heavy quarks, the longitudinal 
momentum fraction dependence in $\phi$ is quite different 
for the heavy-heavy, heavy-light and light-light mesons.  

For the light-light mesons, such as pions, rhos, kaons etc., 
the wavefunction $\phi_{q\overline{q}}
(x,\kappa_\bot)$ must vanish at the endpoint $x=0$ or $1$.  
This can be seen from the kinetic energy term in 
eq.(\ref{lfbse}), where $M_0^2 = {\kappa_\bot^2 + m_1^2 \over x}
- {\kappa_\bot^2 + m_2^2 \over 1-x}$ for the valence Fock space.
To ensure that the bound state equation is well defined
in the entire range of momentum space, $|\phi_{q\overline{q}}
(x,\kappa_\bot)|^2$ must fall down to zero in the
longitudinal direction not slower than $1/x$
and $1/(1-x)$ when $x \rightarrow 0$ and $1$, respectively.
In other words, at least $\phi_{q\overline{q}}(x,\kappa_\bot)
 \sim \sqrt{x(1-x)}$ .
For heavy-light quark mesons, namely the $B$ and $D$ mesons, 
the wavefunction $\phi_{Q\overline{q}}(y,\kappa_\bot)$ is 
required to vanish at $y=0$, where $y$ is the residual 
longitudinal momentum fraction carried by the light quark. 
This is because the kinetic energy in eq.(\ref{Qqbse}) 
only contains a singularity at $y=0$.
On the other hand, since  $0 \leq y \leq \infty$, 
$\phi_{Q\overline{q}}(y,\kappa_\bot)$ should also vanish when 
$y \rightarrow \infty$.  Hence, a possible simple solution is
$\phi_{Q\overline{q}}(y,\kappa_\bot) \sim \sqrt{y}e^{-\alpha y}$ 
or $  \sqrt{y} e^{-\alpha y^2}$.  For heavy quarkonia, 
$-\infty < y < \infty$, the normalization 
forces $\phi_{Q\overline{Q}}(y,\kappa_\bot)$ to vanish as
$y \rightarrow \pm \infty$. Thus, a simple solution may be 
$\phi_{Q\overline{Q}}(y,\kappa_\bot) \sim e^{-\alpha y^2}$.

On the other hand, the transverse momentum dependence in 
these light-front wavefunctions should be more or less
similar.  They all vanish at $\kappa_\bot 
\rightarrow \pm \infty$. A simple form of the $\kappa_\bot$
dependence for these wavefunctions is a Gaussian function:
$e^{- \kappa_\bot^2 / 2\omega^2}$. 

The above analysis of light-front wavefunctions is only 
based on the kinetic energy properties of the constituents.
Currently, many investigations on the hadronic structures use 
phenomenological light-front wavefunctions. One of such 
wavefunctions that has been widely used in the study of 
heavy hadron structure is the BSW wavefunction [\cite{BSW}],
\begin{equation}	\label{bsw}
	\phi_{BSW}(x,\kappa_\bot) = {\cal N} \sqrt{x(1-x)}
		~\exp\left(-{\kappa^2_\bot\over2\omega^2}\right)
		~\exp\left[-{M_H^2\over2\omega^2}(x-x_0)^2\right],
\end{equation}
where ${\cal N}$ is a normalization constant, $\omega$  
a parameter of order $\Lambda_{QCD}$, $x_0=({1\over2} 
-{m_1^2-m_2^2\over2M_H^2})$, and $M_H$, $m_1$, and $m_2$ are 
the hadron, quark, and antiquark masses respectively. In the 
heavy mass limit, the BSW wavefunction can be produced from 
our analysis based on the light-front bound state equations.

Explicitly, for heavy-light quark systems, such as the $B$ and 
$D$ mesons, one can easily find that in the heavy mass limit,
$m_1 = m_Q \sim M_H$, $m_q << m_Q$  so that $x_0 =0$ .
Meanwhile, we also have $M_H x = 
\overline{\Lambda} y$. Furthermore, the factor $\sqrt{x(1-x)}$ 
can be rewritten by $\sqrt{y}$ in according to the corresponding
bound state equation discussed above. Thus, the BSW wavefunction 
is reduced to
\begin{equation}	\label{Qqtwf}
	\phi_{Q\overline{q}}(y,\kappa_\bot) = {\cal N} \sqrt{y}
		~\exp\left(-{\kappa_\bot^2\over2\omega^2}\right)
	~\exp\left(-{\overline{\Lambda}^2\over2\omega^2}y^2\right).
\end{equation}
This agrees with our qualitative analysis given above. Indeed, using 
such a wavefunction we  have already computed the universal 
Isgur-Wise function in $B \rightarrow D, D^*$ decays  [\cite{zhang95}]: 
\begin{equation}
	\xi( v \cdot v') = {1 \over v \cdot v'},
\end{equation}
and from which we 
obtained the slope of $\xi(v \cdot v')$ at the zero-recoil point, 
$\rho^2 = - \xi'(1) = 1$, in excellent agreement with the recent 
CLCO result  [\cite{clco}] of $\rho^2 = 1.01 \pm 0.15 \pm 0.09$. 

For heavy quarkonia, such as the $b\overline{b}$ and $c\overline{c}$
states, $m_1 = m_2 = m_Q$ which leads to $x_0=1/2$ in eq.(\ref{bsw}). 
Also note that $M_H (x- 1/2) = \overline{\Lambda} y$, and the factor 
$\sqrt{x(1-x)}$ must be totally dropped as 
we have discussed form the quarkonium bound state equation. 
Then the BSW wavefunction for quarkonia is reduced to 
\begin{equation}	\label{QQtwf1}
	\phi_{Q\overline{Q}}(y,\kappa_\bot) = {\cal N} ~\exp \left(
		-{\kappa^2_\bot\over2\omega^2}\right) ~\exp\left(
		-{\overline{\Lambda}^2\over2\omega^2}y^2\right),
\end{equation}
which is a form as we expected from the qualitative analysis. 
Here we have not taken the limit of $m_Q \rightarrow \infty$ 
for heavy quarkonia. Thus a possible $m_Q$ dependence in 
wavefunction may be hidden in the parameter $\omega$.

Using the variational approach with the above trial wave function,
we can analytically solve the bound state equation Eq.(\ref{QQbse}).
Under the consideration of SRG invariant for the binding energy
$\overline{\Lambda}$, we find that the effective coupling constant
 [\cite{zhang96}]:
\begin{equation}
	\alpha_\lambda ={g^2_\lambda \over 4 \pi}
		={\pi \over C_f}~ \Bigg( {\overline{\Lambda}^2
		\over \lambda^2}\Bigg)~{1\over a + b \ln {\lambda^2
		\over \overline{\Lambda}^2}}, \label{running}
\end{equation}
where the coefficients $a$ and $b$ are determined by minimizing the
binding energy from (\ref{QQbse}) with (\ref{QQtwf1}). The coefficient 
$b$ is almost a constant (with a weak dependence on $m_Q$ 
but independence on $\overline{\Lambda}$ and $\lambda$), while $a$
depends on both $\overline{\Lambda}$ and $m_Q$, and also slightly on
$\lambda$.  For $\lambda \geq 0.6$ GeV, the $\lambda$-dependence
in the parameter $a$ is negligible.  It is known that 
$\overline{\Lambda}$ is of the same order as  $\Lambda_{QCD}$
which is about $100 \sim 400$ MeV. The charmed 
and bottom quark masses used here are $m_c=1.4$ GeV and 
$m_b=4.8$ GeV.  From the particle data  [\cite{PDB}], the binding 
energy $\overline{\Lambda}$ should also be less than 400 MeV.
For a qualitative consideration, we may take $\overline{\Lambda}
=0.2$ GeV and $\overline{\Lambda}=0.4$ GeV for charmonium. Then
we find that $b=1.15$, and  $a=-0.25$ and $1.1$ respectively.

Numerically, with $\overline{\Lambda} = 200$ MeV and $\lambda = 1$ GeV,
we obtain that
\begin{equation}
	\alpha_\lambda =
	   \left\{ \begin{array}{ll} 0.02665~~~~~~~
		& {\rm charmonium,} \\
		0.06795~~~~ & {\rm bottomonium,} \end{array} \right.
\end{equation}
which is much smaller than that extrapolated from the 
canonical running coupling constant in the naive perturbative 
QCD calculation. 
In order to see how this weak coupling constant varies with the
scale $\lambda$, we take $\overline{\Lambda} =200$ MeV and
vary the value of $\lambda$ around 1 GeV. We find that the 
coupling constant is decreased
very faster with increasing $\lambda$.  In other words, {\it with 
a suitable choice of the hadronic mass scale $\lambda$
in SRG, we can make the
effective coupling constant $\alpha_\lambda$  in $H_\lambda$ 
arbitrarily small. Then the WCT of
nonperturbative QCD can be achieved in terms of $H_\lambda$ 
such that the corrections from $H_{\lambda I}$ can be truly
computed perturbatively}.  This provides the first realization
of the WCT to nonperturbative QCD dynamics on the light-front.

\subsection{Perspectives} 

The applications of the present theory to heavy quarkonium 
spectroscopy and various heavy quarkonium annihilation and 
production processes can be simply achieved by numerically 
solving the bound state equations (\ref{QQbse}), and by further 
including the $1/m_Q$ corrections (which naturally leads to 
the spin splitting interactions). The extension of the 
computations to heavy-light quark systems is straightforward. 
The extension of the present work to light-light hadrons requires 
the understanding of chiral symmetry breaking in QCD which is 
a new challenge to nonperturbative QCD on the light-front.  
Nevertheless, we have provided a detailed analysis to the 
weak-coupling treatment of nonperturbative QCD proposed recently 
 [\cite{Wilson94}]. I believe that   LFQCD 
opens a new research direction in the attempt of solving the most 
difficult problem in field theory, that is, the problem of the 
relativistic composite particle bound states governed by the
nonperturbative dynamics of QCD.

\subsubsection{Acknowledgements:}

I thank A. Harindranath, C. Y. Cheung and G. L. Lin for fruitful 
collaboration. I also thank J. W. Qiu and J. Vary, Darwin Chang 
and C. Q. Geng and H. Y. Cheng for their hospitality at IITAP of 
Iowa State University, National Tsing Hua University and Academia
Sinica, respectively, where the notes is written. Finally, I 
would like to thank the organizers for this stimulating school.

%
%

\end{document}